\documentclass[twocolumn,appendixfloats]{aastex631}

\shorttitle{Data-Driven Radiative MHD Simulations}
\shortauthors{Chen et al.}
\usepackage{amsmath}
\usepackage{graphicx,subfigure}
\usepackage[percent]{overpic}
\usepackage{color}
\hypersetup{linkcolor=red,citecolor=blue}
\newcommand{\sectref}[1]{Section\,\ref{#1}}

\newcommand{\figref}[1]{Figure\,\ref{#1}}
\newcommand{\figsref}[1]{Figures\,\ref{#1}}
\newcommand{\tabref}[1]{Table\,\ref{#1}}

\newcommand{\appdref}[1]{Appendix\,\ref{#1}}

\begin{document}
\title{Data-Driven Radiative Magnetohydrodynamics Simulations with the MURaM code}

\author[0000-0002-1963-5319]{Feng Chen}
\affiliation{School of Astronomy and Space Science, Nanjing University, Nanjing, 210023, China, chenfeng@nju.edu.cn}
\affiliation{Key Laboratory for Modern Astronomy and Astrophysics (Nanjing University), Ministry of Education, Nanjing, 210023, China}
\affiliation{Laboratory for Atmospheric and Space Physics, University of Colorado Boulder, Boulder, CO 80303, USA}

\author[0000-0003-2110-9753]{Mark C. M. Cheung}
\affiliation{CSIRO, Marsfield, NSW 2122, Australia}

\author[0000-0001-5850-3119]{Matthias Rempel}
\affiliation{High Altitude Observatory, National Center for Atmospheric Research, Boulder, CO 80307, USA}

\author[0000-0002-1253-8882]{Georgios Chintzoglou}
\affiliation{Lockheed Martin Solar \& Astrophysics Laboratory, Palo Alto, CA 94304, USA}
\affiliation{University Corporation for Atmospheric Research, Boulder, CO 80307-3000, USA}

\correspondingauthor{Feng Chen}
\email{chenfeng@nju.edu.cn}

\begin{abstract}
We present a method of conducting data-driven simulations of solar active regions and flux emergence with the MURaM radiative magnetohydrodynamics (MHD) code. The horizontal electric field derived from the full velocity and magnetic vectors, is implemented at the photospheric (bottom) boundary to drive the induction equation. The energy equation accounts for thermal conduction along magnetic fields, optically-thin radiative loss, and heating of coronal plasma by viscous and resistive dissipation, which allows for a realistic presentation of the thermodynamic properties of coronal plasma that are key to predicting the observational features of solar active regions and eruptions. To validate the method, the photospheric data from a comprehensive radiative MHD simulation of solar eruption (the ground truth) are used to drive a series of numerical experiments. The data-driven simulation reproduces the accumulation of free magnetic energy over the course of flux emergence in the ground truth with an error of 3\%. The onset time is approximately 8\,min delayed compared to the ground truth. However, a precursor-like signature can be identified at the correct onset time. The data-driven simulation captures key eruption-related emission features and plasma dynamics of the ground truth flare over a wide temperature span from $\log_{10}T{=}4.5$ to $\log_{10}T{>}8$. The evolution of the flare and coronal mass ejection as seen in synthetic extreme ultraviolet images is also reproduced with high fidelity. The method helps to understand the evolution of magnetic field in a more realistic coronal environment and to link the magnetic structures to observable diagnostics.
\end{abstract}

\section{Introduction}\label{sec:intro}
The magnetic field is the essential driver of solar activity. Knowledge of the structure and energy of the coronal magnetic field plays a vital role in understanding the trigger and evolution of major solar eruptions such as flares and coronal mass ejections \citep{Low:1996,Chen:2011}. Due to the difficulty of direct measurement of the magnetic field in the corona \citep{Lin+al:2004}, extensive studies have been performed to model the three-dimensional structure of the coronal magnetic field and its evolution by observation of the magnetic field in the photosphere. The assumption of a vanishing Lorentz force leads to forcefree models\citep{Wiegelmann+Sakurai:2012}, among which nonlinear forcefree field (NLFFF) models are widely used to investigate the magnetic field topology and the energy storage and release during an eruptive event \citep{Schrijver+al:2008}. In addition to solving the forcefree field as a static boundary problem, the magnetofrictional (MF) \citep{Yang+al:1986} method was also used to obtain an NLFFF constrained by the input magnetic field on the boundary. The magnetic field is evolved by assuming that the inductive velocity is proportional to the Lorentz force and eventually relaxes to a forcefree state.  As it is known that the solar atmosphere below the corona is not entirely forcefree \citep{Metcalf+al:1995}, magnetohydrostatic extrapolation \citep{Zhu+Wiegelmann:2018}, which accounts for the static equilibrium between the Lorentz force and plasma forces has been developed to improve the modeling of, for example, the chromospheric magnetic field\citep{Vissers+al:2022}. Magnetohydrodynamic models \citep{Mikic+al:1999} consider the effect of the plasma in a more self-consistent manner and allow for forced magnetic field modeling.

The static magnetic field based on an input magnetogram is regarded as a ``data-constrained" model. This can be extended to a collection of independent solutions obtained from a time series of input magnetic fields \citep[e.g., ][]{Thalmann+Wiegelmann:2008,Amari+al:2014}. The so-called ``data-driven" model solves the continuous temporal evolution of the magnetic field in the domain in response to the change in the magnetic field at the boundary and can be done with MF \citep[e.g., ][]{Cheung+DeRosa:2012,Pomoell+al:2019} or MHD \citep[e.g., ][]{WuST+al:2006,Inoue+al:2018}. The data-driven method has been used to investigate, for example, successive small eruptions \citep{Kaneko+al:2021} and jets \citep{Cheung+al:2015}; emergence \citep{Cheung+DeRosa:2012}, evolution \citep{Hayashi+al:2018,Hayashi+al:2019}, and disposal\citep{Mackay+al:2011} of active regions; and global coronal magnetic field \citep{Fisher+al:2015,Weinzierl+al:2016,Hoeksema+al:2020,Hayashi+al:2022,FengXueshang+al:2023}, as well as the heliospheric impact of CMEs \citep{Jin+al:2018}. In recent years, data-driven simulations have become a powerful tool to model the evolution of flare-productive active regions \citep{Toriumi+WangHaimin:2019,Toriumi:2022} as has been done by \citep{Gibb+al:2014,JiangChaowei+al:2016nc,JiangChaowei+al:2016,Price+al:2019,Price+al:2020,HeWen+al:2020,Klipua+al:2021,Yardley+al:2021,Lumme+al:2022}. As a method that better connects observations and models \citep{Janvier+al:2015}, data-driven simulations also help to reveal the trigger and evolution of eruptions and their driving forces \citep{Inoue+al:2018,LiuChengao+al:2019,GuoYang+al:2019,ZhongZe+al:2021}. A more comprehensive review of data-driven simulations was given recently by \citet{JiangChaowei+al:2022}.

Although data-driven simulations are a successful and significant step forward to bring models closer to actual observations, the role of plasma has been omitted (in MF models) or treated in a rather simplified manner (in most MHD models). Therefore, the comparison between models and observations is done mostly between selected magnetic structures in models and emission features seen in remote sensing images. Realistic coronal models by radiative MHD simulations that take into account the radiative and conductive energy transport for coronal plasma, along the line of works by \citep{Gudiksen+Nordlund:2005a,Gudiksen+Nordlund:2005b}, would be needed, in order to make self-consistent and quantitative model-observation comparisons. \citet{Bourdin+al:2013,Warnecke+Peter:2019} presented realistic models of the active region corona that are constrained/driven by observation-based boundaries and reproduce observed EUV loops. However, these models are for stable (compared with those with ongoing flux emergence or decaying active regions) and nonerupting active regions and are evolved for a short time period. Such realistic simulations of the active region corona are usually computationally expensive, particularly when they encounter the low density area above strong sunspots (high Alfv\'en speed) and extreme plasma temperatures (large thermal diffusivity) during solar flares that will severely limit the timestep of the simulations. An alternative approach is, as done by \citet{Hoeksema+al:2020}, combining MF simulations of long evolution of the global-scale magnetic field and sophisticated radiative MHD simulations \citep{Abbett:2007} that give a more accurate solution for the plasma properties and synthetic observables in the region and time period of interest.

Comprehensive radiative MHD simulations of solar flares \citep{Cheung+al:2019} and the emergence of large flare-productive active regions \citep{Chen+al:2022} are made possible by the extension of the MURaM, which can cope with the aforementioned extreme conditions. In this paper, we present the development of the MURaM code to enable data-driven realistic simulations of solar active regions and flares. These simulations may help to quantify the properties of plasma that evolve consistently with the magnetic field. Not only may these simulations allow for a more accurate calculation of the magnetic field, if plasma forces ever matter, but they also help to quantitatively predict the consequences of eruptions, such as flare class, which has never been done. The remainder of the paper is organized as follows. We describe the numerical method of the data-driven MURaM code, in particular the implementation of the boundary condition, in \sectref{sec:method} and present in \sectref{sec:res} a comparison between the magnetic field and plasma properties of the ground truth data and the radiative MHD simulation driven by the photospheric observable from the ground truth. Control experiments are discussed in \sectref{sec:a_exp}. We summarize the results and conclude in \sectref{sec:sum}.

\section{Numerical Method}\label{sec:method}
We adapt the MURaM code with the coronal extension \citep{Voegler+al:2005,Rempel:2017} to conduct simulations driven by photospheric boundaries. The code solves the evolution of mass, momentum, plasma energy and magnetic field in a conservative treatment. The equations are
\begin{align}
   \frac{\partial \rho}{\partial t} = &-\nabla\cdot\left(\rho\mathbf{v}\right)\\
   \frac{\partial \rho\mathbf{v}}{\partial t} = &-\nabla\cdot\left(\rho\mathbf{v}\mathbf{v}+p\mathbf{I}\right) + \rho \mathbf{g} + \mathbf{F}_{\rm L} + \mathbf{F}_{\rm SR} \\
   \frac{\partial E_{\rm hd}}{\partial t} = &-\nabla\cdot\left(\mathbf{v}(E_{\rm hd}+p)+q\mathbf{\hat{b}}\right) \nonumber\\
                                                                    &+ \mathbf{v}\cdot\left( \rho \mathbf{g} + \mathbf{F}_{\rm L} + \mathbf{F}_{\rm SR}\right) + Q_{\rm loss} + Q_{\rm h}\\
   \frac{\partial q}{\partial t} = &\frac{1}{\tau_{q}}\left(-f_{\rm sat}\kappa_{\parallel}\mathbf{\hat{b}}\cdot\nabla T - q\right)                           \\
   \frac{\partial \mathbf{B}}{\partial t} = &\nabla\cdot\left(\mathbf{v}\mathbf{B}-\mathbf{B}\mathbf{v} \right).                                       
\end{align}
The energy conservation is solved for the plasma energy $E_{\rm hd}{=}\rho e_{\rm int} + \frac{1}{2}\rho \mathbf{v}^2$. $\mathbf{F}_{\rm SR}$ represents the Boris correction \citep{Boris:1970:BC} term that imposes an artificially reduced light speed ($c$), which limits the maximum Alfv\'en velocity allowed in the system and follows the expression
\begin{align}
\mathbf{F}_{\rm SR} = & -(1-f_{A})[\mathbf{I}-\mathbf{\hat{b}}\mathbf{\hat{b}}](-\rho (\mathbf{v}\cdot\nabla)\mathbf{v} \nonumber\\
                                    & -\nabla p + \rho \mathbf{g} + \mathbf{F}_{\rm L}),
\end{align} 
as derived by \citet{Rempel:2017} following \citet{Gombosi:etal:2002:SR}.
The Lorentz force $\mathbf{F}_{\rm L}$ is given by 
\begin{align}
\mathbf{F}_{\rm L} =& f_{A}\frac{1}{4\pi}\nabla\cdot\left(\mathbf{B}\mathbf{B}-\frac{1}{2}\mathbf{I}\mathbf{B}^2\right) \nonumber \\
                                &+(1-f_{A})\frac{1}{4\pi}(\nabla\times\mathbf{B})\times\mathbf{B},
\end{align}
which follows a conservative treatment in high $\beta$ regions ($f_{A} \to 1$) as other flux terms and is preserved perpendicular to $\mathbf{B}$ in low $\beta$ regions.
The factor $f_{A}$ determines the Alfv\'en velocity limitation and is chosen as 
\begin{equation}
f_{A} = \frac{1}{\sqrt{1+\left(\frac{v_{A}}{c}\right)^4}},
\end{equation}
such that the corrected Alfv\'en velocity ($v_{c}$) following the expression
\begin{equation}
v^2_{c} = \frac{v^2_{A}}{\sqrt{1+\left(\frac{v_{A}}{c}\right)^4}}
\end{equation}
changes smoothly between the uncorrected and corrected regime. The heat conduction along magnetic field lines $\mathbf{q}$ is solved by an auxiliary equation, such that $\mathbf{q}$ approaches the Spitzer heat conduction \citep{Spitzer:1962} on a timescale $\tau_{q}$ and saturates on the level limited by $f_{\rm sat}$. $Q_{\rm loss}$ is the total energy loss through optically thin radiation provided by the CHIANTI \citep{Landi+al:2012}. The radiative transfer of optically thick radiation is not considered in the data-driven version of the MURaM code. $Q_{\rm h}$ is a simple Newtonian-type background heating given by 
\begin{equation}
Q_{\rm h} = \rho(\overline{e_{0}}(z) - e_{\rm int})/\tau_{\rm h},
\end{equation}
where $\tau_{\rm h}=10^{3}$\,s controls the time scale on which this term (only acts as heating) brings the temperature to that of the horizontally averaged initial quiet Sun stratification $\overline{e_{0}}(z)$. In practice, the contribution of this artificial heating is very small, as most of the active region corona in the data-driven simulations is hotter than the target temperature (presented in the results) and evolves dynamically on a shorter time scale. The plasma heating in the corona is mainly provided by viscous and resistive dissipation ($Q_{\rm vis}$ and $Q_{\rm res}$), as described in \citep{Rempel:2017}. These dissipation terms are not written in the equation above but have been taken into account (implicitly or explicitly) in the energy equation, such that energy conservation is preserved, as shown in \citep{Chen+al:2022}. In contrast to typical MURaM simulations where the energy flux of coronal heating is generated by granular motion braiding field lines, the energy flux in the data-driven version is provided through boundary driving as done in this study or implemented by \citet{Fan:2022}.  A detailed description of the formulation of the Lorentz force, Boris correction, hyperbolic heat conduction and optically thin radiative loss has been presented by \citet{Rempel:2017}. The equation is solved with the numerical scheme described by \citet{Rempel:2014}.

\subsection{Implementation of the data-driven boundary}\label{sec:method_imp}
\begin{figure}
\includegraphics[width=9cm]{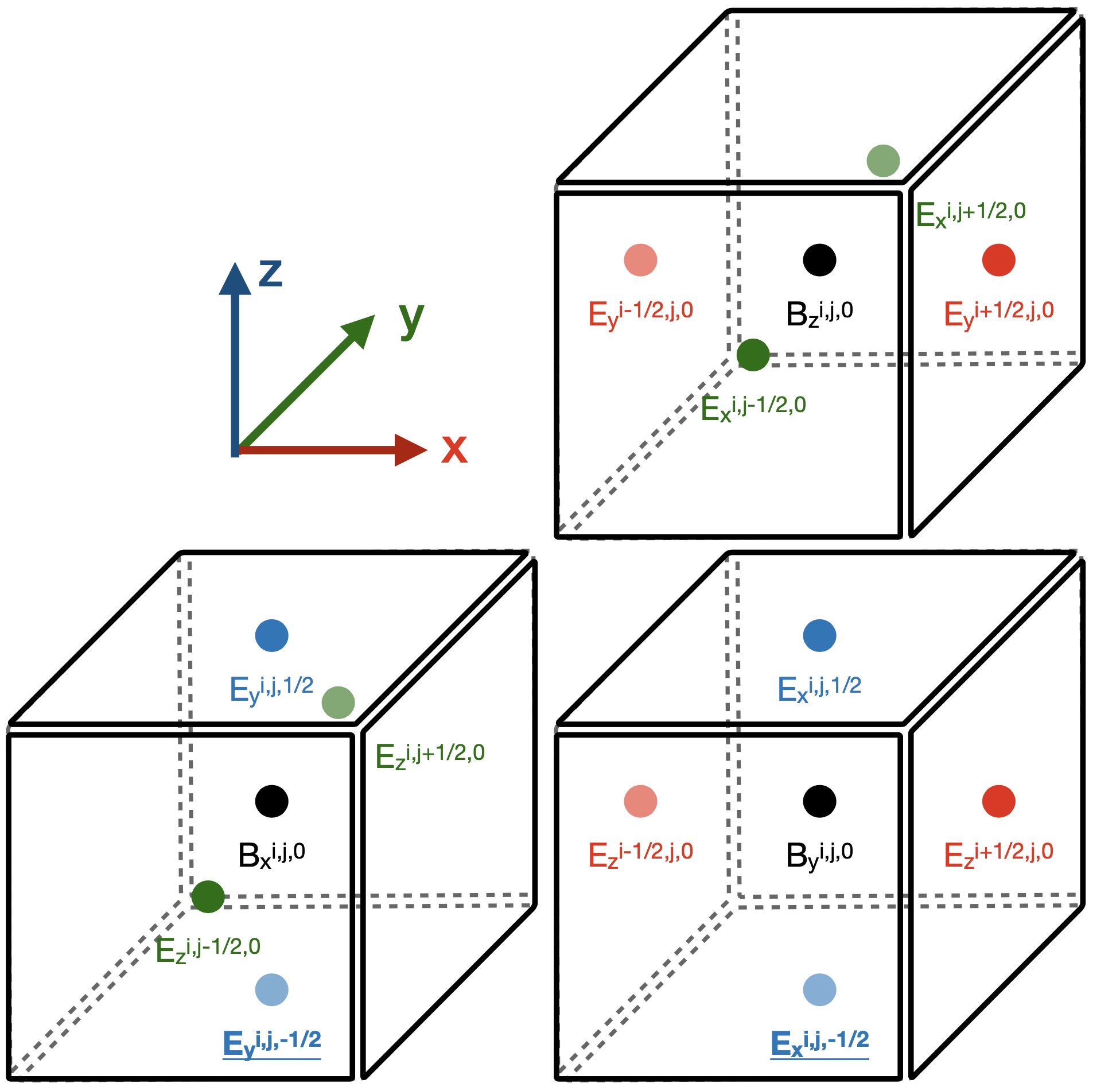}
\caption{Schematic diagram of the induction equation in the first layer cells in the computation domain (k=0) shown in Equations \ref{equ:bz}, \ref{equ:bx}, and \ref{equ:by}. The bottom boundary is at $k{=}-1/2$. Quantities in far-side surfaces of the cells ($i-1/2$, $j+1/2$, and $k=-1/2$) are displayed with a slightly higher transparency. The underscored quantities correspond to the driver electric field implemented at the bottom boundary. 
\label{fig:induction}} 
\end{figure}

\begin{figure*}
\includegraphics{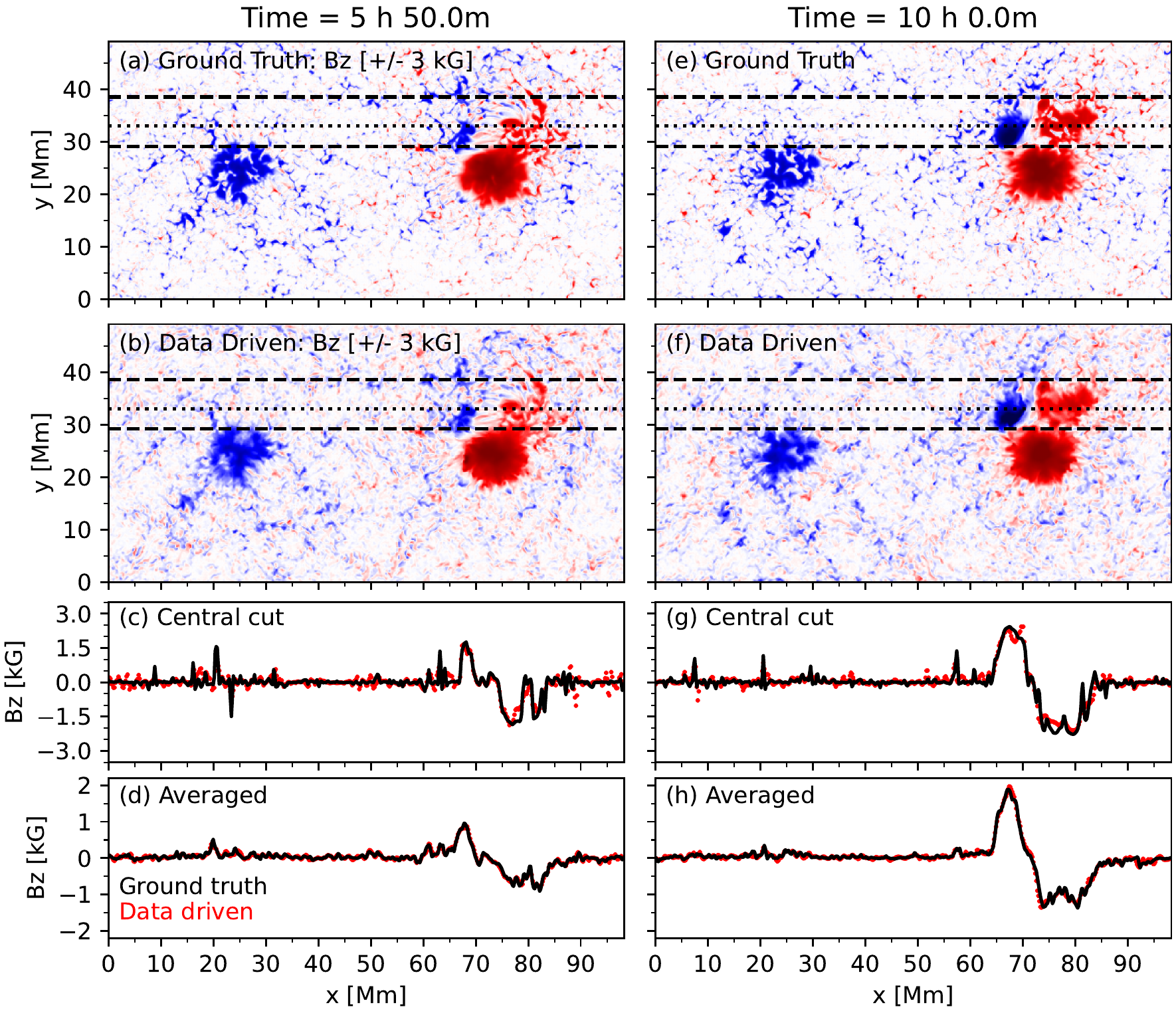}
\caption{Validation of the data-driven boundary. (a): $B_{z}$ extracted from the photospheric height of the ground truth simulation. (b): $B_{z}$ at the first layer in the computational domain of the data-driven boundary test with $\Delta z{=}64$\,km. (c): A comparison along the dotted line in the magnetogram of $B_{z}$ in the ground truth simulation (black) and that in the data-driven boundary test (red). (d) : $B_{z}$ in the ground truth and the data-driven boundary test averaged between the two dashed lines in the magnetogram. (e), (f), (g), and (h): The result near the end of the evolution shown in the same fashion as in Panels (a), (b), (c), and (d).
\label{fig:compare_bz}} 
\end{figure*}

The most essential new ingredient in the data-driven module of MURaM is the time-dependent bottom boundary that drives the evolution of the magnetic field in the computational domain. All physical quantities in MURaM are defined at cell centers as 
$$U^{i,j,k},$$
where $i, j,$ and k are integers and are evolved by
\begin{align}
\frac{\Delta U^{i,j,k}}{\Delta t} &= \frac{1}{\Delta x}\left(F_{x}^{i+\frac{1}{2},j,k} - F_{x}^{i-\frac{1}{2},j,k}\right) \nonumber\\
                                                 &+\frac{1}{\Delta y}\left(F_{y}^{i,j+\frac{1}{2},k} - F_{y}^{i,j-\frac{1}{2},k}\right)  \nonumber\\
                                                 &+\frac{1}{\Delta z}\left(F_{z}^{i,j,k+\frac{1}{2}} - F_{z}^{i,j,k-\frac{1}{2}}\right),
\end{align}
where $F_{x}$, $F_{y}$, and $F_{z}$ are fluxes at cell interfaces.
The data-driven boundary is implemented at the bottom face of the first layer cells in the computational domain (i.e., $F_{z}^{i,j,-\frac{1}{2}}$). 
We first revisit the discretized induction equation of $B_{z}$ in the first layer cells ($k{=}0$), which reads
\begin{align}\label{equ:bz}
\frac{\Delta B_{z}^{i,j,0}}{\Delta t} = &-\frac{1}{\Delta x}\left(E_{y}^{i+\frac{1}{2},j,0} - E_{y}^{i-\frac{1}{2},j,0}\right) \nonumber\\
                                                        &+\frac{1}{\Delta y}\left(E_{x}^{i,j+\frac{1}{2},0} - E_{x}^{i,j-\frac{1}{2},0}\right),
\end{align}
where components in the $\mathbf{v}\mathbf{B}-\mathbf{B}\mathbf{v}$ tensor have been rewritten as electric fields at the (side) interfaces of the cells. This reveals a simple fact that $B_{z}$ will not be evolved by a vertical flux, i.e, the flux through the $k{\pm}\frac{1}{2}$ faces. See also the illustration in \figref{fig:induction}.

The induction equations of the horizontal components of the magnetic field are written as
\begin{align}\label{equ:bx}
\frac{\Delta B_{x}^{i,j,0}}{\Delta t} = & -\frac{1}{\Delta y}\left(E_{z}^{i,j+\frac{1}{2},0} - E_{z}^{i,j-\frac{1}{2},0}\right) \nonumber\\
                                                        &+\frac{1}{\Delta z}\left(E_{y}^{i,j,\frac{1}{2}} - \underline{E_{y}^{i,j,-\frac{1}{2}}}\right)
\end{align}
\begin{align}\label{equ:by}
\frac{\Delta B_{y}^{i,j,0}}{\Delta t} = &\frac{1}{\Delta x}\left(E_{z}^{i+\frac{1}{2},j,0} - E_{z}^{i-\frac{1}{2},j,0}\right) \nonumber\\
                                                        -&\frac{1}{\Delta z}\left(E_{x}^{i,j,\frac{1}{2}} - \underline{E_{x}^{i,j,-\frac{1}{2}}}\right).
\end{align}
It is clear that $B_{x}^{i,j,0}$ ($B_{y}^{i,j,0}$) can be changed by a difference of $E_{y}$ ($E_{x}$) between the top and bottom faces of the first layer cells, as also illustrated in \figref{fig:induction}. The underscored terms in Equations \ref{equ:bx} and \ref{equ:by} correspond to the driver electric fields. Moreover, when evaluating$ E_{y}^{i,j,\frac{1}{2}}$ and $E_{x}^{i,j,\frac{1}{2}}$, we replace the standard fourth-order four-point stencil used in MURaM with a second order two-point stencil such that these fluxes are determined by values inside the domain. To summarize, the fluxes through the side and top faces of a first layer cell are determined by quantities within the computational domain, and only the flux through the bottom face (i.e., the intended data-driven boundary) contains information from external data.

\subsection{Deriving the electric field as the boundary driver}\label{sec:method_efield}
In this study, we use the electric field derived from the photospheric magnetic and velocity fields of the ground truth simulation \citep{Cheung+al:2019} to drive a data-driven simulation and reproduce the flare that occurs in that simulation. The velocity field ($\mathbf{v}^{\rm pho}$) and magnetic field ($\mathbf{B}^{\rm pho}$) are extracted from a horizontal layer ($k{=}116$) that corresponds to the photosphere level in the cube of the ground truth simulation at approximately 60\,s cadence. An electric field following the ideal Ohm's law is given by
\begin{equation}
\mathbf{E}_{h}^{vb} = \left(-\mathbf{v}^{\rm pho}\times\mathbf{B}^{\rm pho}\right)_{h},
\end{equation}
where the subscript indicates that only the horizontal components are considered. Due to the presence of numerical diffusion and limited cadence of sampling (60\,s vs. timestep of ~0.01\,s), the difference in consecutive snapshots of the vertical magnetic field ($\Delta B^{\rm pho}_{z}$) is not exactly equal to the time integral of $\nabla_{h}\times\mathbf{E}^{vb}_{h}$ over $\Delta t$. Therefore, to ensure that the derived electric field reproduces the evolution of $B^{\rm pho}_{z}$, a correction electric field $\mathbf{E}^{c}_{h}$ is determined by solving
\begin{align}
\nabla_{h}\times \mathbf{E}^{c} &= -\frac{\Delta B^{\rm pho}_{z}}{\Delta t} - \nabla_{h}\times \mathbf{E}^{vb}_{h} \\
\nabla_{h}\cdot \mathbf{E}^{c} &= 0.
\end{align}
Following \citet{Cheung+DeRosa:2012}, $E_{c}$ is solved with Fourier method with periodic horizontal boundary, which is also consistent with the horizontal boundary used in the simulations presented in this study.
Finally, the electric field that is used to drive the simulation is given by
\begin{equation}
\mathbf{E}_{h} = \mathbf{E}^{vb}_{h} + \mathbf{E}^{c}_{h}.
\end{equation}
For the time series of the extracted velocity and magnetic field, $\mathbf{E}_{h}$ is evaluated for each time interval, which means that the electric field at each grid point is a piecewise constant function of time.

We note that if the cadence of the input magnetic field is significantly reduced (e.g., 720\,s instead of 60\,s used in this study), there is a noticeable increase in spurious high spatial frequency oscillations in $E_{c}$, due to generally larger changes between consecutive magnetograms. This can be largely eliminated by rebinning the low cadence magnetograms to a lower spatial resolution. Nevertheless, we can still perform a data-driven simulation in the original or even higher resolution with rebinned input magnetic field, from which the electric field is derived, by the strategy described later in \sectref{sec:a_resolution}.

We also note that the conservative treatment (and non-staggered mesh) of the numerical scheme used in this code makes the implementation of the driver electric field different from other codes that implement electric field driving on a staggered mesh \citep[e.g.,][]{Cheung+DeRosa:2012,Hayashi+al:2018,Pomoell+al:2019,Inoue+al:2022}. An alternative approach is using a velocity field derived from observations (sometimes in combination of imposing a time series of magnetogram), as done by, for example, \cite{GuoYang+al:2019}, \cite{HeWen+al:2020}, \cite{Kaneko+al:2021}, and \cite{WangXinyi+al:2023}. \cite{Hayashi+al:2019} proposed a method of using both electric and velocity fields. In comparison, we do not constrain velocity field at the boundary by prescribed values, and the dynamics is determined by in-domain evolution subject to the electric field driving.

\subsection{Validation of the data-driven boundary}\label{sec:res_bound}
To validate the effect of the data-driven boundary, we conduct a series of tests using a simplified zero-$\beta$ fashion method. We omit the plasma forces and thermodynamics and consider only the plasma motion driven by the Lorentz force and the induction equation. The equations solved read
\begin{align}
   \frac{\partial \rho_{s}\mathbf{v}}{\partial t} = &-\nabla\cdot\left(\rho_{s}\mathbf{v}\mathbf{v}\right) + \mathbf{F_{L}} + \mathbf{F_{SR}} \\
   \frac{\partial \mathbf{B}}{\partial t} = &\nabla\cdot\left(\mathbf{v}\mathbf{B}-\mathbf{B}\mathbf{v} \right),                
\end{align}
where $\rho_{s}{=}3\times10^{-7}$\,g cm$^{-3}$ is the static background density. The Boris correction is still used in order to be consistent with production runs described in \sectref{sec:method_setup}. It also helps to handle higher Alfv\'en velocity above sunspots and improve time steps, although given the relatively large $\rho_{s}$, Alfv\'en velocity is no longer a severe issue in this boundary test. The simulation domain is a thin slab that has a large horizontal extent of $L_{x}\times L_{y}{=}98.304\times49.152$\,Mm$^{2}$ but spans only $L_{z}=2.048$\,Mm in the vertical direction. The domain is resolved by $N_{x}\times N_{y}\times N_{z}{=}512\times256\times32$ gridpoints, resulting in $\Delta x{=}\Delta y{=}192$\,km and $\Delta z{=}64$\,km resolutions in the horizontal and vertical directions, respectively. The bottom boundary is symmetric for magnetic vectors and horizontal velocities but anti-symmetric (i.e., close) for vertical mass flux to prevent the photosphere from running away. The top boundary is open/close for out-/in-flows. Lateral boundaries are periodic for all quantities. The boundary test starts from $t{=}0$\,min and is evolved for 10 hours, which allows us to examine how well the data-driven boundary reproduces not only the emergence of the parasitic bipole (beginning at $t{\approx}260$ in the photosphere) but also the evolution of small-scale magnetic features under magnetoconvection.

A comparison between the ground truth magnetogram and $B_{z}$ (first layer in the computational domain) in the data-driven boundary test is presented in \figref{fig:compare_bz}. The chosen snapshots correspond to the middle of the fast emergence stage (left column) and almost the end of the simulation (right column). Although magnetic structures in the data-driven $B_{z}$ appear to be slightly smoother compared with the ground truth, they retain the stable main sunspot pair, follow the emerging parasitic bipole, and resemble smaller flux concentrations in the quiet Sun region. For the interest of this study, we focus in particular on the region of the emerging parasitic bipole. Along a slice through the center of the parasitic bipole, the data-driven $B_{z}$ quantitatively reproduces the emergence of the bipole. There are some noticeable mismatches in small-scale magnetic features. This is partly introduced by the electric fields as they sample the evolution of the ground truth magnetic field with a finite cadence and partly because of the limited vertical resolution. However, if we examine the $B_{z}$ averaged over the $y$-extent of the parasitic bipole (between the dashed lines in \figref{fig:compare_bz}), the differences on the smallest scales are eliminated (somehow indicating their random nature), and the average profiles of the ground truth and data-driven test become consistent. A discussion of how the boundary test may be further improved is presented in \sectref{sec:a_boundary}.

\subsection{Simulation Setup}\label{sec:method_setup}
\begin{table*}
\caption{Summary of simulation cases\label{tab:list}}
\center
\begin{tabular}{l|cccccccccl}
\hline
\hline
Run Name & Resolution$^{3}$ & $t_{0}$/$t_{\rm peak}$ & Class & $\Delta E_{\rm free}$ $^4$  & $\epsilon_{f}$ & $C_{\rm vec}$ & $C_{\rm CS}$& $E^{\prime}_{\rm n}$ $^5$ & $E^{\prime}_{\rm m}$ $^5$ & Shown in\\
~ & $N_{x}{\times}\Delta x$ [km] & [min] & ~ & [$10^{30}$\,erg] & ~ & ~ & ~ & ~ & ~ & Figure\\
\hline
Ground~truth~evo. & ~\,$512{\times}192$  & 0 / 472 & -- & 4.38 & 1 & 1 & 1 & 1 & 1 & \ref{fig:vstime}\\
Ground~truth~flare & ~\,$512{\times}192$  & 445 / 474 & C4.3 & 4.46 & - & - & - & - & - & \ref{fig:vstime},\ref{fig:em},\ref{fig:aia},\ref{fig:vapor},\ref{fig:vlos_full},\ref{fig:flaretime_lines},\ref{fig:em_full}\\
\hline
512\_t250 (ref.)$^{1}$ & ~\,$512{\times}192$  & 250 / 487 & -- & 4.32 & 0.97 & 0.95 & 0.97 & 0.78 & 0.77 & \ref{fig:vstime},\ref{fig:vstime_appd},\ref{fig:flaretime}\\
512\_t250\_flare (ref.)$^{1}$ & ~\,$512{\times}192$  & 452 / 483 & C2.6 & 4.65 & - & - & - & - & - & \ref{fig:vstime},\ref{fig:em},\ref{fig:aia},\ref{fig:vapor},\ref{fig:vlos_full},\ref{fig:vstime_appd},\ref{fig:flaretime_lines},\ref{fig:em_full}\\
512\_t250\_v3000$^{2}$ & ~\,$512{\times}192$  & 250 / 487 & C1.9 & 3.52 & 0.96 & 0.95 & 0.97 & 0.77 & 0.77 & \ref{fig:vstime},\ref{fig:flaretime},\ref{fig:em-full-512-t250-v3000}\\
512\_t250\_v1500$^{2}$ & ~\,$512{\times}192$  & 250 / 481 & C1.8 & 4.56  & 0.96 & 0.95 & 0.97 & 0.77 & 0.77 & \ref{fig:vstime},\ref{fig:flaretime},\ref{fig:em-full-512-t250-v1500}\\
512\_t250\_v300 & ~\,$512{\times}192$  & 250 / 482 & -- & 4.08 & 0.97 & 0.95 & 0.97 & 0.77 & 0.77 & \ref{fig:vstime},\ref{fig:flaretime}\\
512\_t250\_v300\_flare & ~\,$512{\times}192$  & 452 / 483 & C2.9 & 4.14 & - & - & - & - & - & \ref{fig:vstime},\ref{fig:em-full-512-t250-v300-flare}\\
\hline
512\_t200 & ~\,$512{\times}192$  & 200 / 483 & -- & 4.67  & 0.96 & 0.95 & 0.97 & 0.77 & 0.76 & \ref{fig:vstime_appd}\\
512\_t200\_flare & ~\,$512{\times}192$  & 453 / 484 & C5.8 & 5.51 & - & - & - & - & - & \ref{fig:vstime_appd},\ref{fig:em-full-512-t200-flare}\\
512\_t250\_rerun1 & ~\,$512{\times}192$  & 250 / 496 & -- & 5.05 & 0.98& 0.95 & 0.97 & 0.77 & 0.77 & \ref{fig:flaretime}\\
512\_t250\_rerun2 & ~\,$512{\times}192$  & 250 / 490 & -- & 5.21 & 0.97 & 0.95 & 0.97 & 0.78 & 0.77 & \ref{fig:flaretime}\\
512\_t250\_rerun3 & ~\,$512{\times}192$  & 250 / 513 & -- & 6.81 & 0.97 & 0.95 & 0.97 & 0.77 & 0.77 & \ref{fig:flaretime}\\
512\_t250\_v500 & ~\,$512{\times}192$  & 250 / 482 & -- & 2.35 & 0.97  & 0.95 & 0.97 & 0.77 & 0.76 & \ref{fig:flaretime}\\
512\_t144 & ~\,$512{\times}192$  & 144 / 517 & -- & 7.59  & 0.96 & 0.94 & 0.97 & 0.76 & 0.76& \ref{fig:vstime_appd}\\
\hline
256\_t250 & ~\,$256{\times}384$  & 250 / 501 & -- & 6.97  & 0.89 & 0.96 & 0.97 & 0.79 & 0.78& \ref{fig:vstime_appd}\\
256\_t250\_flare & ~\,$256{\times}384$  & 470 / 501 & C3.3 & 6.52 & - & - & - & - & - &  \ref{fig:vstime_appd},\ref{fig:em-full-256-t250-flare}\\
256\_t200 & ~\,$256{\times}384$  & 200 / 515 & -- & 6.73  & 0.87 & 0.96 & 0.97 & 0.79 & 0.78 & \ref{fig:vstime_appd}\\
256\_t200\_flare & ~\,$256{\times}384$  & 470 / 514 & C2.9 & 6.82 & - & - & - & - & - & \ref{fig:vstime_appd},\ref{fig:em-full-256-t200-flare}\\
256\_t144 & ~\,$256{\times}384$  & 144 / 518 & -- & 7.30  & 0.87 & 0.95 & 0.97 & 0.77 & 0.76& \ref{fig:vstime_appd}\\
\hline
128\_t250 & ~\,$128{\times}768$  & 250 / 511 & -- & 9.86 & 0.80 & 0.96 & 0.98 & 0.80 & 0.80& \ref{fig:vstime_appd}\\
128\_t250\_flare & ~\,$128{\times}768$  & 470 / 513 & C5.7 & 9.34 & - & - & - & - & - & \ref{fig:vstime_appd},\ref{fig:em-full-128-t250-flare}\\
128\_t200 & ~\,$128{\times}768$  & 200 / 522 & -- & 8.93  & 0.77 & 0.96 & 0.97 & 0.8 & 0.79& \ref{fig:vstime_appd}\\
128\_t200\_flare & ~\,$128{\times}768$  & 489 / 521 & C3.0 & 8.63 & - & - & - & - & - & \ref{fig:vstime_appd},\ref{fig:em-full-128-t200-flare}\\
128\_t250\_shift & ~\,$128{\times}768$  & 250 / 507 & -- & 9.66  & 0.80 & 0.97 & 0.98 & 0.8 & 0.79 & \ref{fig:vstime_appd}\\
128\_t250\_shift\_flare & ~\,$128{\times}768$  & 470 / 508 & C8.5 & 9.36 & - & - & - & - & - & \ref{fig:vstime_appd},\ref{fig:em-full-128-t250-shift-flare}\\
128\_t250\_rebin & ~\,$512{\times}192$  & 250 / 510 & -- & 6.49  & 0.91 & 0.95 & 0.97 & 0.78 & 0.78 & \ref{fig:vstime_appd}\\
128\_t250\_rebin\_flare & ~\,$512{\times}192$  & 470 / 508 & C1.5 & 4.58  & - & - & - & - & - & \ref{fig:vstime_appd},\ref{fig:em-full-128-t250-rebin-flare}\\
\hline
\end{tabular}
\tablecomments{
\begin{enumerate}
\item Reference data-driven evolution and flare runs.
\item Evolution and flare runs done in one go. 
\item For all runs: $N_{y}{=}N_{x}/2$ and $\Delta y{=}\Delta x$;  $N_{z}=768$ and $\Delta z{=}64$\,km.
\item $\Delta E_{\rm free}$ is determined by the difference of the pre-flare maximum and the post-flare minimum within a time window of about 30\,min centered at $t_{\rm peak}$.
\item $E^{\prime}_{\rm n (m)}=1-E_{\rm n (m)}$
\end{enumerate}
}
\end{table*}

The simulation domain, which is identical to the ground truth simulation in size, covers an area of $L_{x}\times L_{y}{=}98.304\times49.152$\,Mm$^2$, with a vertical extent of $L_{z}{=}49.152$\,Mm. The vertical domain starts from the photosphere, whereas that of the ground truth simulation starts from about 8\,Mm beneath the photosphere. Thus, the data-driven simulation reaches a slightly higher height. The domain is resolved by a mesh of $N_{x}\times N_{y}\times N_{z}{=}512\times256\times768$, yielding 192\,km and 64\,km grid spacing in the horizontal and vertical directions, respectively. 

\begin{figure}
\includegraphics{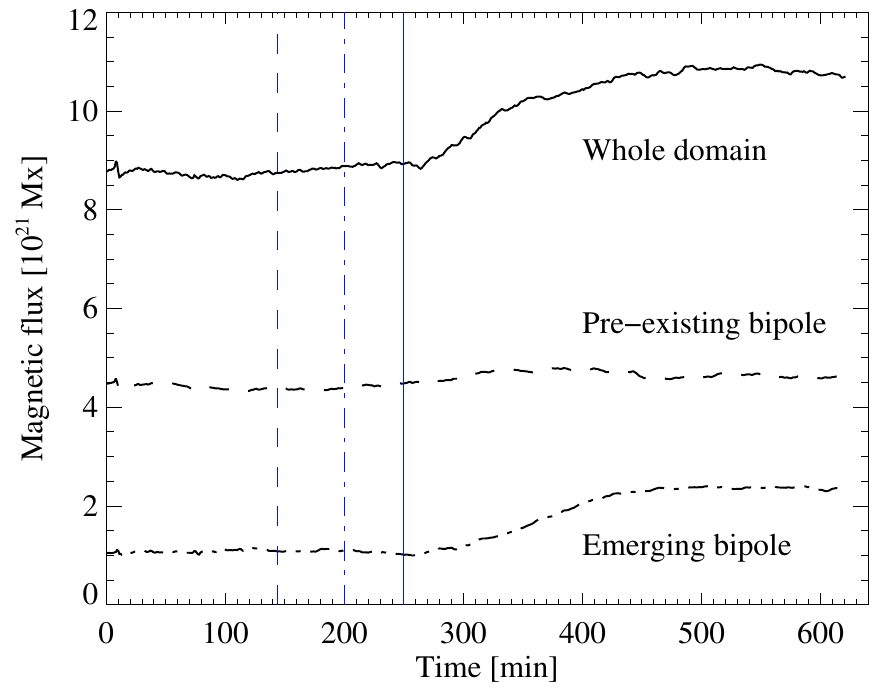}
\caption{Unsigned magnetic flux in the photosphere. Fluxes are also calculated for the $y$ range of the pre-existing bipole and emerging parasitic bipole. The blue dashed, dash-dotted, and solid lines mark the starting times of ``t144", ``t200", and ``t250" runs, respectively.
\label{fig:starttime}}
\end{figure}

The initial condition of the data-driven simulation is constructed as follows. We adopt a cube of $256\times256\times768$ from a quiet Sun simulation that has a horizontal domain of $256\times256$ (periodic boundary and $\Delta x{=}\Delta y{=}192$\,km) with a sufficiently large vertical domain of 1536 grid points (116 grid points beneath photosphere and $\Delta z{=}64$\,km). The quiet Sun simulation is fully relaxed in the sense that the small-scale dynamo in the convective layers maintains a mixed-polarity magnetic field and the quiet Sun magnetic field provides a sufficient energy flux to maintain a million K corona. Then, the extracted cube is copied in the $x$ direction to form a cube of $512\times256\times768$. With periodic horizontal boundaries and the assumption that $B_{z}$ vanishes at infinity\footnote{The net vertical flux is zero in the ground truth simulation}, we calculate the potential field from the photospheric $B_{z}$ of the ground truth simulation at $t_{0}$ and add the potential field to the existing mixed-polarity magnetic field in the quiet Sun cube. For the reference run, we choose $t_{0}{=}250$\,min, which is shortly before the parasitic bipole starts to emerge in the photosphere, as shown in \figref{fig:starttime}. Velocities in the region of $|B|{\ge}1500$\,G, is set to zero to avoid unrealistic disturbance inside sunspots, while the remainder of the cube retains the velocity field of the quiet Sun cube.

The horizontal boundaries are periodic for all quantities. The top boundary is open for outflows and closed for inflows of mass flux. The magnetic field at the top boundary is a potential field calculated from $B_{z}$ at the uppermost cell in the domain.  The vertical mass flux at the bottom is anti-symmetric (closed) to prevent runaway solutions. The bottom boundary is symmetric for magnetic vectors and horizontal momentum. The time series of the electric field derived from the ground truth simulation is implemented in the induction equation as described in \sectref{sec:method_efield}. The density and specific internal energy in the ghost cells at the bottom boundary are set to the mean values in the photosphere of the quiet Sun cube that is used to construct the initial snapshot via a factor $f=0.05$, i.e., 
\begin{equation}
\rho_{t+\Delta t}^{\rm ghost} = (1-f)~\rho_{t}^{\rm ghost} + f~\overline{\rho_{0}}|_{z=0},
\end{equation}
which drives these quantities to the prescribed mean values at a timescale of $1/f$ iterations.

The ground truth simulation was performed with two setups. An ``evolution run" simulates the whole evolution (${\approx}10$\,hrs) of the emergence of a parasitic bipole into a stable bipole active region, with upper limits on the plasma velocity (3000\,km s$^{-1}$) and temperature ($5\times10^{7}$\,K). A ``flare run", which is the setup presented in \citet{Cheung+al:2019}, is restarted from a snapshot of the long run with virtually no limit on plasma velocity and temperature to fully capture the violent dynamics ( maximum $|\mathbf{v}|{\approx}$3600\,km s$^{-1}$) and hot plasma ($T>10^{8}$\,K) during the flare. Another experiment performed at the same time showed that the setup with velocity  (300\,km s$^{-1}$) and temperature ($10^{7}$\,K) ceilings used for typical non-eruptive active regions is sufficient for pre-flare evolution. This motivates us to test different setups to explore a more optimistic strategy that could combine consistency of results and efficiency of resources.

We perform a data-driven simulation (``512\_t250\_v3000" in \tabref{tab:list}) with the identical velocity limit used by the ground truth evolution run, but \emph{no} limit on temperature as done in the ground truth flare run. This run captures the dynamics and temperature during the flare in one go, as the maximum velocity in this data-driven run is about $2800$\,km  s$^{-1}$. In this paper, we focus on runs performed with a two-stage strategy as follows. The reference data-driven run, ``512\_t250" in \tabref{tab:list}, is an evolution run with $\mathbf{v}_{\rm max}{=}200$km s$^{-1}$ and $T_{\rm max}{=}10$\,MK. It is evolved from $t_{0}{=}250$\,min to the end of the ground truth simulation (about $t{=}620$\,min). We evaluate the flux of 1--8 \AA~soft X-ray according to the temperature response of GOES-15 satellite. Although the flux given by the evolution run is not quantitatively meaningful as high temperatures are suppressed, the time of the peak $t_{\rm peak}$ is used to determine the initial time of the flare rerun. The reference data-driven flare run, ``512\_t250\_flare" in \tabref{tab:list}, is started from about 30\,min before $t_{\rm peak}$ of the evolution run, with no limit on velocity and temperature as done in the ground truth flare run.

Extensive control experiments on resolutions,  start time $t_{0}$, parameters and initial condition are performed, as discussed in \sectref{sec:a_exp}. Flare runs are also performed for (some of) the control experiments. A complete list of runs conducted is given in \tabref{tab:list}. Hereafter, the data-driven (flare) simulation refers to the reference (flare) run, if no specific run name is given.

\section{Results}\label{sec:res}
\subsection{General properties of the magnetic free energy and flare}\label{sec:res_mag}

\begin{figure*}
\includegraphics{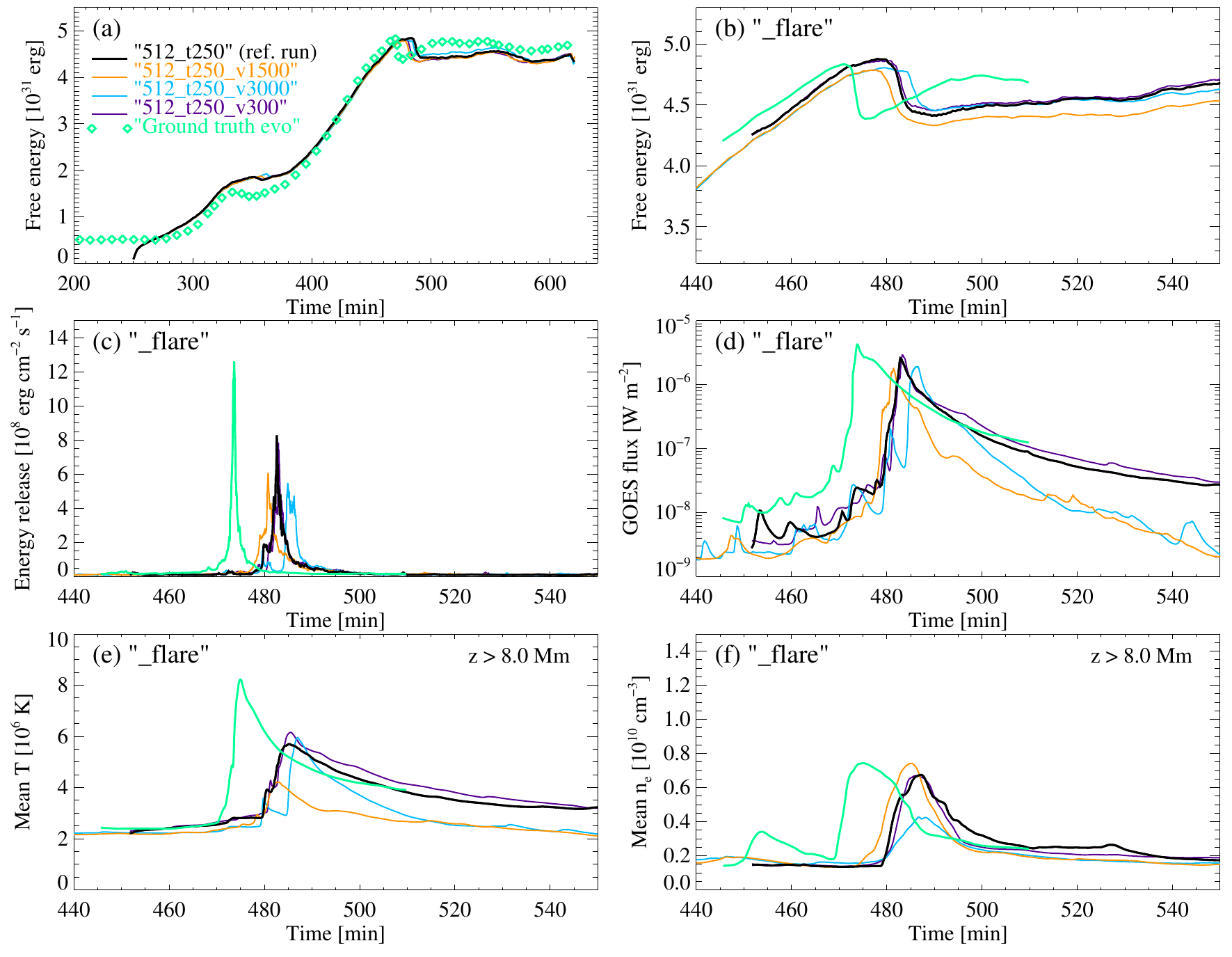}
\caption{General properties of the simulation runs listed in \tabref{tab:list}. (a): Free magnetic energy for evolution runs. The corresponding run name is marked after each combination of line style and color, which also applies to all following panels in this figure. See \sectref{sec:method_setup} for details. (b): Free magnetic energy in the flare runs. (c): Magnetic energy release rate as the sum of Lorentz force work and (numerical) resistive heating. (d): Synthetic {\it GOES} 1--8\AA~flux. (e) \& (f): Mean temperature and electron number density in the coronal domain higher than  8\,Mm.
\label{fig:vstime}}
\end{figure*}

We plot in \figref{fig:vstime}(a) the temporal evolution of free magnetic energy $E_{\rm free}$, which is the difference between the energy of the actual magnetic field and the potential field. The potential field is calculated from $B_{z}$ in the first layer with periodic lateral boundaries and vanishing $B_z$ at infinity. The free magnetic energy of the ground truth run is almost unchanged before $t{\approx}250$\,min, after which it increases gradually with time as the parasitic bipole emerges into the photosphere. A small transient event is found at $t{\approx}340$\,min, resulting in a small decay of free magnetic energy. The flare occurs at $t{\approx}470$\,min in the ground truth evolution run, as indicated by a steep drop of the free magnetic energy by about $4\times10^{30}$\,erg. In the data-driven simulation, $E_{\rm free}$ first needs to accumulate from null in the initial condition (potential field). Then, the temporal evolution of the free energy in the data-driven simulation is consistent with that of the ground truth until the flare occurs. The small transient event also occurs in the data-driven simulation, although it is clearly delayed and weaker. This makes the curve of the data-driven run (black line in \figref{fig:vstime}(a)) deviate temporally from that of the ground truth run (green diamonds in \figref{fig:vstime}(a)). However, the two curves converge in the evolution after $t{\approx}380$\,min. 

For a quantitative comparison, we calculate the ratio of the free energy in the data-driven (DD) simulation relative to that in the ground truth (GT) simulation:
\begin{equation}
\epsilon_{f}=\frac{E^{\rm DD}_{\rm free}}{E^{\rm GT}_{\rm free}}.
\end{equation}
The value of $\epsilon_{f}$ at $t{=}460$\,min, which is shortly before the flare occurs, is 0.97. Therefore, the data-driven simulation is able to accurately reproduce the amount of free magnetic energy of the ground truth simulation with an error of ~3\%. Moreover, to evaluate how well the data-driven simulation reproduces the magnetic field in the ground truth simulation in general, we evaluate the following metrics as done by \citet{Schrijver+al:2006}:
\begin{align}
C_{\rm vec} &=\frac{\sum\mathbf{B}_{\rm GT}\cdot\mathbf{B}_{\rm DD}}{\left(\sum|\mathbf{B}_{\rm GT}|^2\sum|\mathbf{B}_{\rm DD}|^2\right)^{1/2}} \\
C_{\rm CS} &=\frac{1}{N}\sum\frac{\mathbf{B}_{\rm GT}\cdot\mathbf{B}_{\rm DD}}{|\mathbf{B}_{\rm GT}||\mathbf{B}_{\rm DD}|} \\
E_{\rm n} &=\frac{\sum|\mathbf{B}_{\rm GT}-\mathbf{B}_{\rm DD}|}{\sum|\mathbf{B}_{\rm GT}|} \\
E_{\rm m} &= \frac{1}{N}\sum\frac{|\mathbf{B}_{\rm GT}-\mathbf{B}_{\rm DD}|}{|\mathbf{B}_{\rm GT}|},
\end{align}
where $\mathbf{B}$ is the magnetic vector in the \emph{common domain} of the ground truth and data-driven simulations, and $N$ is the total number of gridpoints in this domain. 

It is obvious that all metrics yield unity for the ground truth. Larger values of the coefficients indicate a higher similarity between the magnetic fields. The correlation coefficients ($C_{\rm vec}$ and $C_{\rm CS}$) at $t{=}460$\,min suggest a good general match between the data-driven and ground truth simulations. The normalized and mean errors seem to deviate slightly more from unity. These values (0.77 and 0.78) are on a similar level to those achieved by other modeling methods analyzed by \citet{Schrijver+al:2006}. A more comprehensive evaluation of the coefficients and errors, including their temporal evolution and comparison with the performance of nonlinear forcefree models and data-driven MHD models, is an ongoing project and will be presented in a separate study. We perform the same calculation of the metrics and free energy ratio at $t{=}460$\,min for all evolution runs of the control experiments, as listed in \tabref{tab:list}. To compare with lower resolution runs, the ground truth data cube is binned\footnote{Binning is done only in the horizontal direction.} to lower resolutions by averaging. However, $E^{\rm GT}_{\rm free}$ always refers to the value estimated from the cube in the original resolution.

The flare in the data-driven (evolution) run occurs at about $t{=}480$\,min, which can be identified as a sudden decay in the curve of $E_{\rm free}$ similar to that found in the ground truth evolution run. We note again that the flare timing and energetics in evolution runs are only provided for record. The flare properties are only considered quantitatively meaningful in flare reruns shown later in the paper, because the violent plasma dynamics during the eruption will likely affect the release of magnetic energy.  It is interesting to note that at almost the time of the ground truth flare ($t{=}472$\,min), the curve of $E_{\rm free}$ shown in \figref{fig:vstime}(a) tends to bend down, which more clearly shown in \figref{fig:flaretime}. This shape indicates an enhanced dissipation of the free magnetic energy, although it does not overturn the general increasing trend. This behavior is also confirmed in control experiments and implies that a flare precursor is predicted at the correct flare onset time, as discussed in \sectref{sec:a_time},

We start the flare run with a snapshot at $t{=}452$\,min. The pre-eruption evolution of the free energy in the flare run is almost identical to that in the evolution run, and hence, it is also consistent with that of the ground truth, as shown in \figref{fig:vstime}(b). This behavior is found for all evolution-flare run pairs because the flare setup, which allows for higher temperatures and velocities, only makes a difference after the flare occurs. The eruption in the data-driven flare run occurs at about $t{=}480$\,min. $t_{\rm peak}$ is delayed by 8\,min compared to that of the ground truth flare\footnote{The $t_{\rm peak}$ in \tabref{tab:list} is rounded to the nearest integer, thus showing a delay of 9\,min.}. We also note that the time when flare occurs in the ground truth simulation may also be different with changes in parameters and resolutions. Because the flare is slightly delayed, the data-driven simulation accumulates more free energy than the ground truth does (by about $2\times10^{30}$\,erg). The free energy released by the data-driven flare is $4.65\times10^{30}$\,erg, which matches the result of the ground truth flare ($4.46\times10^{30}$\,erg) with an error of 4\%.

We plot in \figref{fig:vstime}(c) the magnetic energy release rate, which is estimated by the sum of total Lorentz force work and resistive dissipation integrated over the coronal volume and converted to a dimension of energy flux when divided by the horizontal area of the domain, as done in \citet{Chen+al:2022}. The pre-eruption energy release rate in the ground truth simulation has a base level of about $1.5\times10^{7}$\,erg cm$^{-2}$ s$^{-1}$. This is the energy source that maintains the temperature and dynamics of coronal plasma, as shown in earlier by \citet{Rempel:2017,Chen+al:2022}. In comparison, the base magnetic energy release rate in the data-driven simulation is about $10^{7}$\,erg cm$^{-2}$ s$^{-1}$. 

The energy release of flares appears as strong peaks that are about one order of magnitude higher than the base level. The ground truth flare demonstrates a clearly higher and narrower peak than the data-driven flare, which can also be inferred from the steeper slope of the $E_{\rm free}$ curve of the ground truth flare in panel (b). It is important to note that because the volume integrated energy release is divided by the total area of the domain when converted to a flux, the peak flux of a flare is obviously underestimated. If we consider the filling factor of the flare-related domain to be 0.01-0.1, the resulting energy flux of a flare will be on the order of $10^{10}$ to $10^{11}$\,erg cm$^{-2}$ s$^{-1}$. This range is similar to that used in flare models that assumes an ejection of energy flux at the flare loop top \citep[e.g., ][]{Allred+al:2015,Reep+al:2015,Kowalski+al:2017,Hong+al:2022}\footnote{These flare models assume a flux of non-thermal electrons, which is not considered in the MURaM simulation. Thus, the comparison is only in the sense that the amplitudes of the fluxes are comparable.}.

As shown in \figref{fig:vstime}(d), the GOES flux of the data-driven flare features a few small peaks between $t{=}450$ and 480\,min, which is caused by smaller transients before the main flare. The timing of these peaks in the data-driven flare run is even consistent with those seen in the GOES flux of the ground truth flare. The main flare gives rise to a steep increase from pre-flare level of $10^{-8}$\,W m$^{-2}$ to the peak value of $2.6\times10^{-6}$\,W m$^{-2}$, corresponding to C2.6 class, which is somehow lower than the C4.3 ground truth flare. The gradual decay after the peak lasts for a considerable time until the simulation is terminated.

\begin{figure*}
\includegraphics{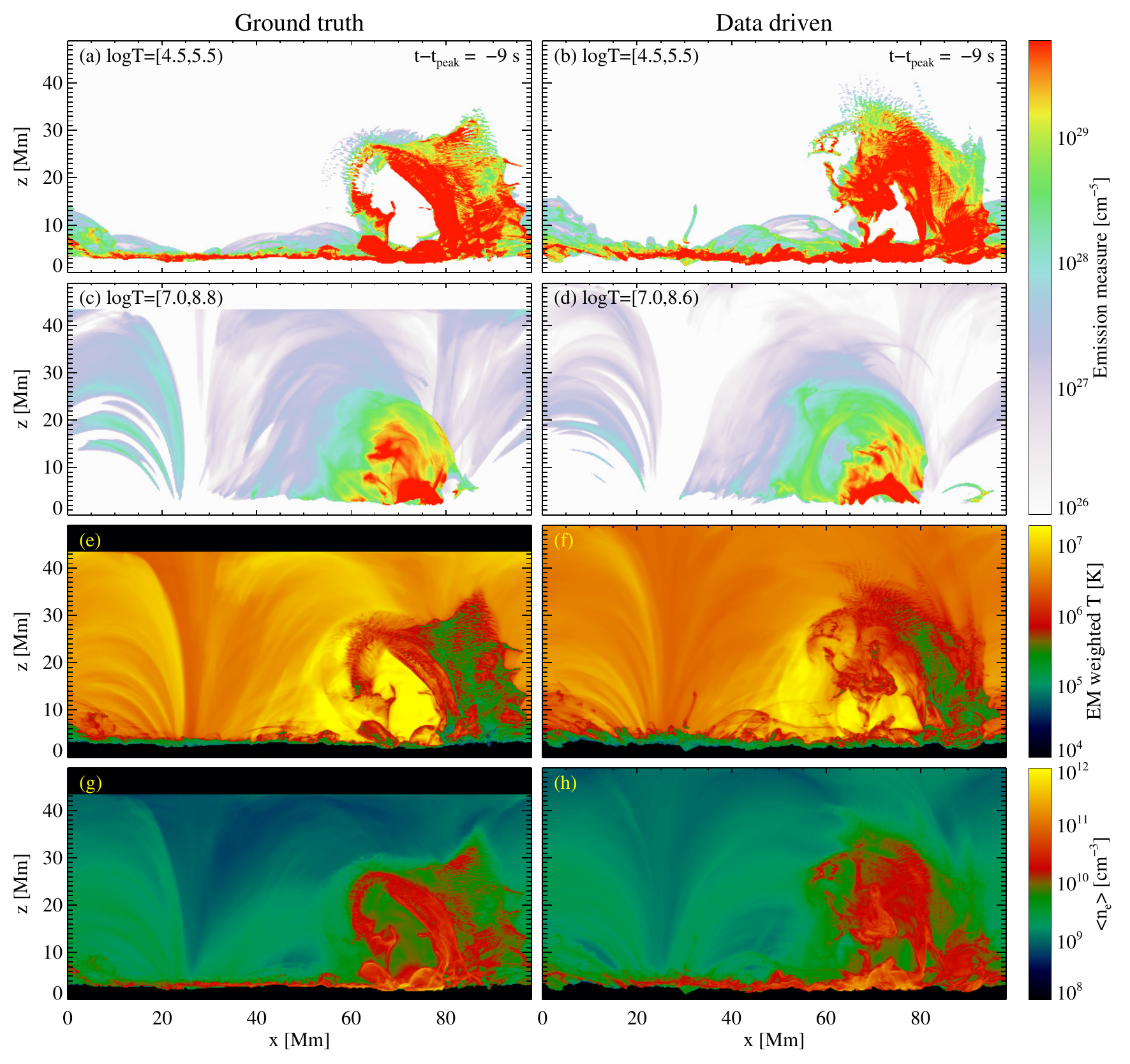}
\caption{Comparison of the emission features in the ground truth (left column) and data-driven (right column) simulations. (a) \& (b): Emission measure from plasma cooler than $10^{5.5}$\,K. (c) \& (d): Emission measure from plasma hotter than $10^{7}$\,K. (e) \& (f): Emission measure weighted temperature from a horizontal view along the $y$ axis. (g) \& (h): Mean number density along the line-of-sight estimated from the emission measure of the full temperature range. An animated version of the figure displays the temporal evolution of 10\,min starting from about 300\,s before the flare peak.
\label{fig:em}}
\end{figure*}

We compare the (arithmetic) mean temperature and number density of the coronal domain above $z{=}8.0$\,Mm in the data-driven and ground truth simulations in \figref{fig:vstime}(e) and (f). The data-driven simulation has a similar mean temperature and density as the ground truth simulation before the flare, which is anticipated as the non-eruptive stage of the data-driven simulation has a comparable energy flux with the ground truth simulation\footnote{More specifically, the temperature ratio $T_{\rm GT}{/}T_{\rm DD}=1.1$ is consistent with the ratio of the flux energy flux $1.5^{2/7}$, following the classic scaling relation of \citet{RTV}. The mean number density is approximately the same, although a slightly lower density (by a factor of $1.5^{4/7}$) is predicted by the scaling relation. This is because the effect of omitting the radiative transfer in the lower atmosphere and adding Newtonian-type heating leads to a thicker chromoshere (by about 1.5\,Mm) in the data-driven simulation and higher density at the coronal base.}.  During the flare, the corona never reaches equilibrium and is highly structured, as will be shown in \figref{fig:em}. Nonetheless, the mean temperature and density present an overall assessment of the strength of the flare. The peak mean temperature in the data-driven flare reaches about 6\,MK; however, it is still considerably lower than the peak mean temperature of the ground truth simulation, which is over 8\,MK. The increase in the mean number density mostly highlights the mass of the coronal mass ejection, since the selected height is higher than the top of hot post-flare loops. The peak mean density in the data-driven simulation is similar to that in the ground truth flare. A spatially-resolved inspect of the temperature and density distribution near the flare peak will be presented in the following section.

\subsection{Emission signatures near the flare peak}\label{sec:res_em}

\begin{figure*}
\includegraphics{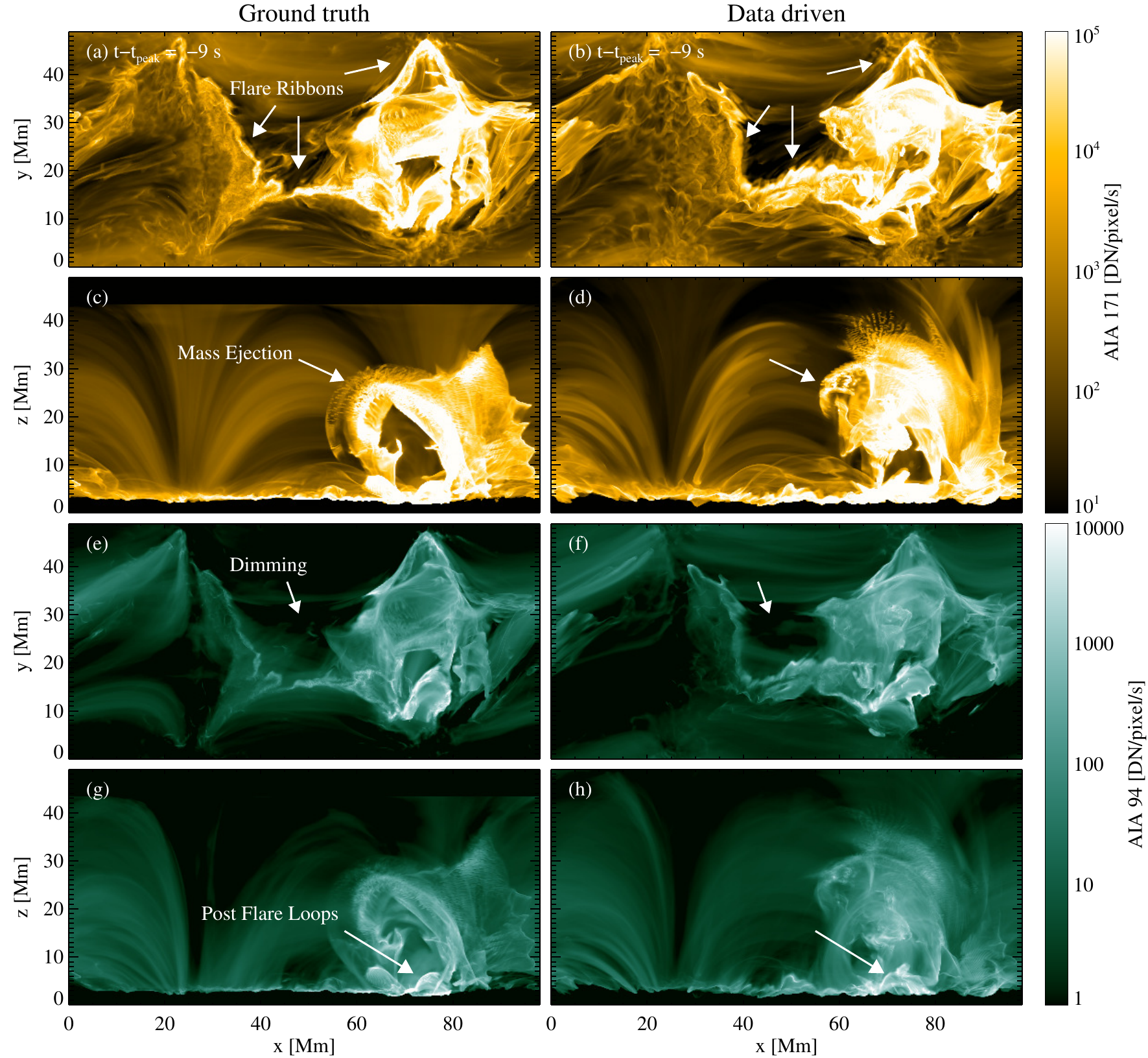}
\caption{Comparison of the emission features in the ground truth (left column) and data-driven (right column) simulations. (a) \& (b): Synthetic AIA 171\AA~ channel images from a top view along the $z$ axis. (c) \& (d): Synthetic AIA 171\AA~ channel images from a horizontal view along the $y$ axis. (e) \& (f): Synthetic AIA 94\AA~ channel images from a top view along the $z$ axis. (g) \& (h): Synthetic AIA 94\AA~ channel images from a horizontal view along the $y$ axis. Arrows indicate some key features of the flare following Supplementary Figure 5 of \citet{Cheung+al:2019}.  An animated version of the figure displays the temporal evolution of 10\,min starting from about 300\,s before the flare peak.
\label{fig:aia}}
\end{figure*}

In this section, we investigate the emission signatures of the plasma involved in the eruption and the magnetic field configuration. In particular, we focus on the snapshot at the same {\emph stage} of the eruption, i.e., 9\,s before the flare peak, when the erupted flux rope is most clearly seen in the middle of the vertical domain.

Panels (a)--(d) of \figref{fig:em} present the side-view emission measure (EM) contributed by cool/hot plasma in the ground truth and data-driven flare simulations. We integrate the differential emission measure (DEM) between $\log_{10}T{=}$4.5 and 5.5 to represent the cool plasma, and to account for the hot plasma, the DEM is integrated for all temperatures higher than 10\,MK. A comparison between the EM through the full temperature ranges of the two simulations is presented in \figref{fig:em_full} in \appdref{sec:a_compare_em}.

The lower-temperature EM of the data-driven simulation in \figref{fig:em}(b) shows a large arch of dense cool plasma reaching over 30\,Mm height. Although the detailed structures are visibly different, the erupted plasma in the data-driven simulation reproduces the general arch-like shape of that in the ground truth simulation shown in \figref{fig:em}(a).  In the high-temperature EM, the data-driven flare gives rise to a series of hot loops between $x{=}50$ and 80\,Mm. The brightest post-flare loops after magnetic reconnection are found lying between $x{=}65$ and 80\,Mm, right above the emerging parasitic bipole. The spatial distribution of the hot plasma EM of the data-driven flare resembles in general that of the ground truth simulation shown in \figref{fig:em}(c). The most discernible difference is a small patch around $(x,z){\approx}(70,15)$, immediately above the dense post-flare loops. The more detailed comparison in \figref{fig:em_full} shows that the plasma in this area the data-driven simulation are mostly in $6.4{<}\log_{10}T{<}7$.

Furthermore, we evaluate the EM weighted mean temperature from the same line-of-sight. Panels (e) and (f) of \figref{fig:em} show similar temperature maps: an low-temperature arch embedded in higher temperatures. The coolest structure, which corresponds to the $10^5$\,K plasma carried by the erupted flux rope, is found along the right leg of the arch (above $x{\approx}$90\,Mm). The left leg of the arch turns out to be hotter ($10^6$\,K), because this part is involved in reconnection with pre-existing magnetic field and is heated, as will be shown later. The highest temperature is found below the erupted flux rope and immediately above the emerging parasitic bipole. High-temperature structures also spread to the left due to thermal conduction along the field lines connected to the reconnection region. 

The animation associated with \figref{fig:em} displays a temporal evolution of about 10\,min around $t_{\rm peak}$ (from 310s before to 250 after $t_{\rm peak}$. Despite the differences in some detailed structures discussed above, the overall kinematic evolution of the mass ejection and hot flare loops are highly similar in both simulations.

A comparison of synthetic EUV images of the AIA 171\AA~ and 94\AA~ channels is presented in \figref{fig:aia}. The snapshot of the ground truth flare is 9\,s earlier than that shown in Supplementary Figure 5 in \citet{Cheung+al:2019} but illustrates the same features. An animation associated with this figure covers the time period as for \figref{fig:em}. The synthetic images of the data-driven flare reproduce most of the key features of the ground truth flare. The AIA 171\AA~ images show multiple flare ribbons in the top view (see arrows in \figref{fig:aia}(a) \& (b)) and the mass ejection in the side view (arrows in \figref{fig:aia}(c) \& (d)), while the AIA 94\AA~ images illustrate the post-flare loops and a coronal dimming region (arrows in \figref{fig:aia}(e) \& (f)). As a side note, \citet{Cheung+al:2019} indicates post-flare loops in the 171\AA~ channel images, but here, we indicate these loops in the 94\AA~ channel images, and they refer to the same structure.

The animation also confirms the consistency between the observables produced by the ground truth and data-driven flare. It is interesting to note that some detailed dynamics can be reproduced in high fidelity, for instance, evolution of the flare ribbon as a growing bright thread (AIA 171\AA~top view) and a series of loops (AIA 94\AA~top view) connecting the flare ribbon to the reconnection region. \citet{Cheung+al:2019} interpreted the flare ribbon as footpoints of magnetic field lines that undergo slipping reconnection. The consistency of the observables indicate that the data-driven flare does reproduces the magnetic connectivity, as well as the reconnection process.

\subsection{Magnetic topology near the flare peak}\label{sec:res_bfield}
\begin{figure*}
\includegraphics{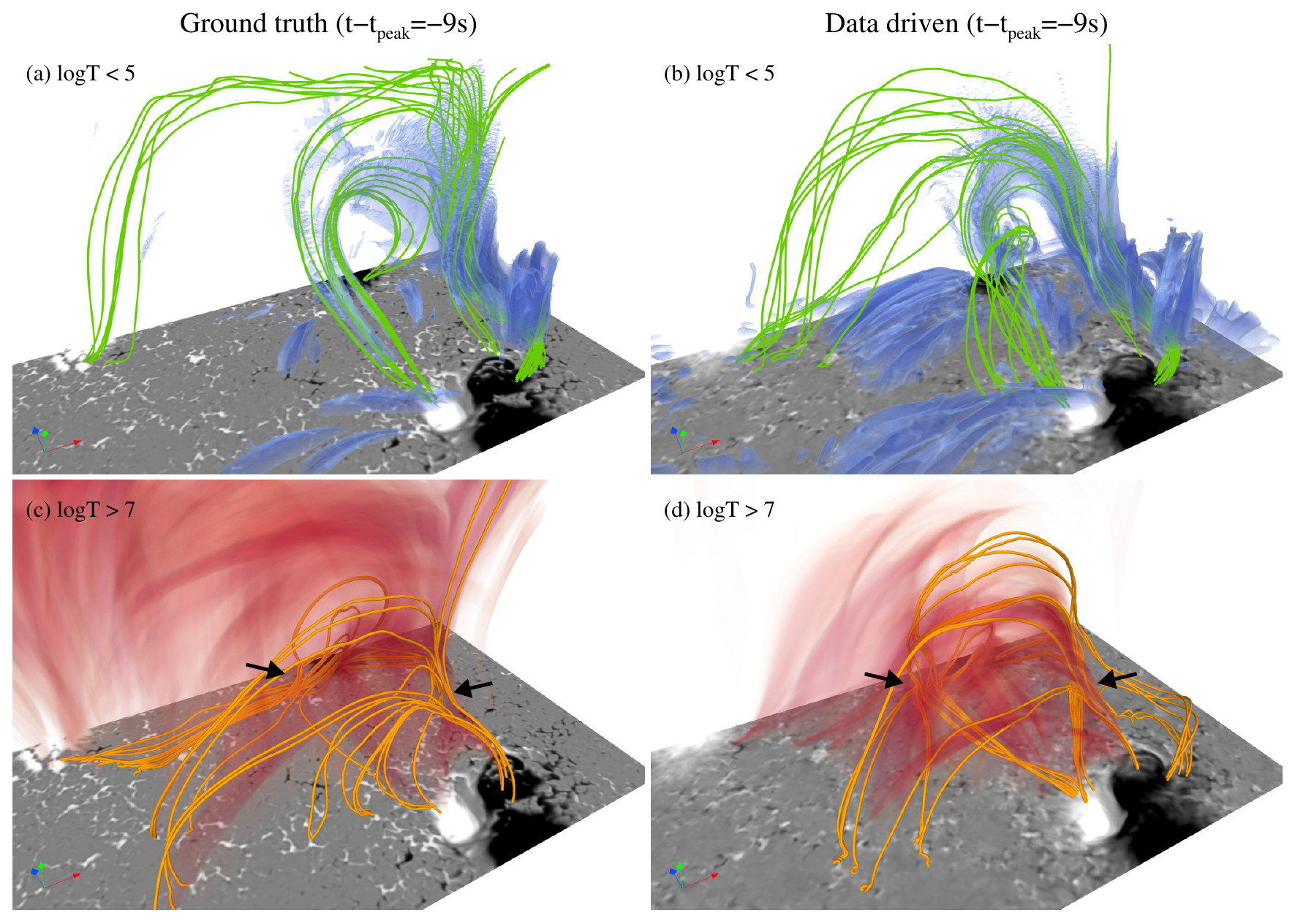}
\caption{A 3D illustration of the magnetic field lines coupled with the cool and hot plasmas. The left and right columns present the ground truth flare and the data-driven flare, respectively. The upper panels display the magnetic field lines (green lines) hosting cool plasma of $\log_{10}T{<}5$ (bluish semi-transparent features). The lower panels show the magnetic field lines (orange lines) passing through regions of hot plasma of $\log_{10}T{>}7$ (reddish semi-transparent features). The arrows indicate reconnection sites, which are discussed in \citet{Cheung+al:2019} and \sectref{sec:res_em} of this paper.
\label{fig:vapor}}
\end{figure*}

We illustrate the magnetic configuration of the magnetic field near the flare peak time in \figref{fig:vapor}. In particular, we focus on the magnetic field lines that elevate cool plasma and those involved in magnetic reconnection and form hot post-flare loops. The opaque features render the square of the electron number density within a certain temperature range, which corresponds to the EM without integration along the line-of-sight. The upper (lower) panels of \figref{fig:vapor} render plasma of $T{<}10^5$\,K ($T{>}10^7$\,K). Seed points that are used to trace magnetic field lines are distributed along the displayed features with a bias to low (high) temperatures. The field of view has been shifted by $L_{y}/2$ as done in \citet{Cheung+al:2019}, such that the eruption is observed in the field-of-view and the pre-existing bipole appears on the boundary. 

As shown in \figref{fig:vapor}(a) and (b), the flux rope magnetic field lines that carry the cool plasma connect the parasitic bipole, where they originate, through an inclined arcade. They also connect from the parasitic negative spot to the pre-existing positive spot through a large arch roughly along the diagonal direction of the domain and connect from the parasitic positive spot to the pre-existing negative spot through a lower arch roughly along the $y$ direction. The connectivity across the domain is given rise by the reconnection process illustrated in Supplementary Figure 4 (a) and (b) of \citet{Cheung+al:2019}. 

Panels (c) and (d) reveal the two reconnection sites before the flare peak, as indicated by the black arrows. One is located near the center of the field of view and is between the parasitic bipole and the pre-existing bipole on the father side. This reconnection gives rise to high-temperature large arches, which are more clearly seen in panel (c). The reconnection also impacts field lines connecting to the small positive magnetic elements in the quiet Sun area (lines that reach the lower left of the field of view of \figref{fig:vapor}) and produces the flare ribbons shown in \figref{fig:aia}. The other reconnection site is located above the parasitic bipole, as illustrated in Supplementary Figure 4 (c) of \citet{Cheung+al:2019}. The magnetic field lines restore the connection between the two spots in the parasitic bipole after the eruption of the flux rope. These short loops over the parasitic bipole are filled by the evaporation of plasma that is heated to over 10\,MK and are the major source of soft X-ray emission during the flare.

\subsection{Plasma dynamics near the flare peak}\label{sec:a_dyn}
\begin{figure*}
\gridline{\fig{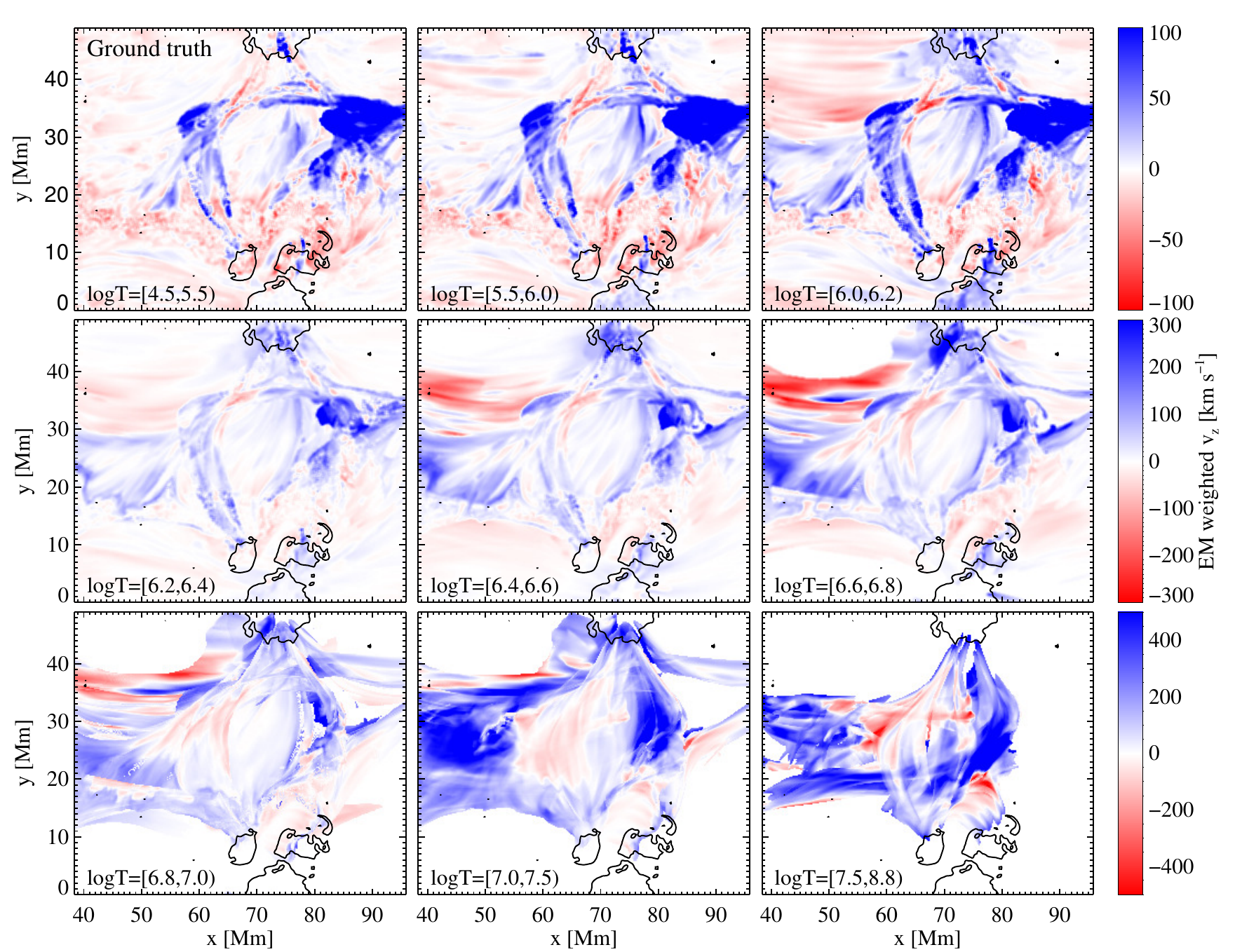}{0.7\textwidth}{}}
\gridline{\fig{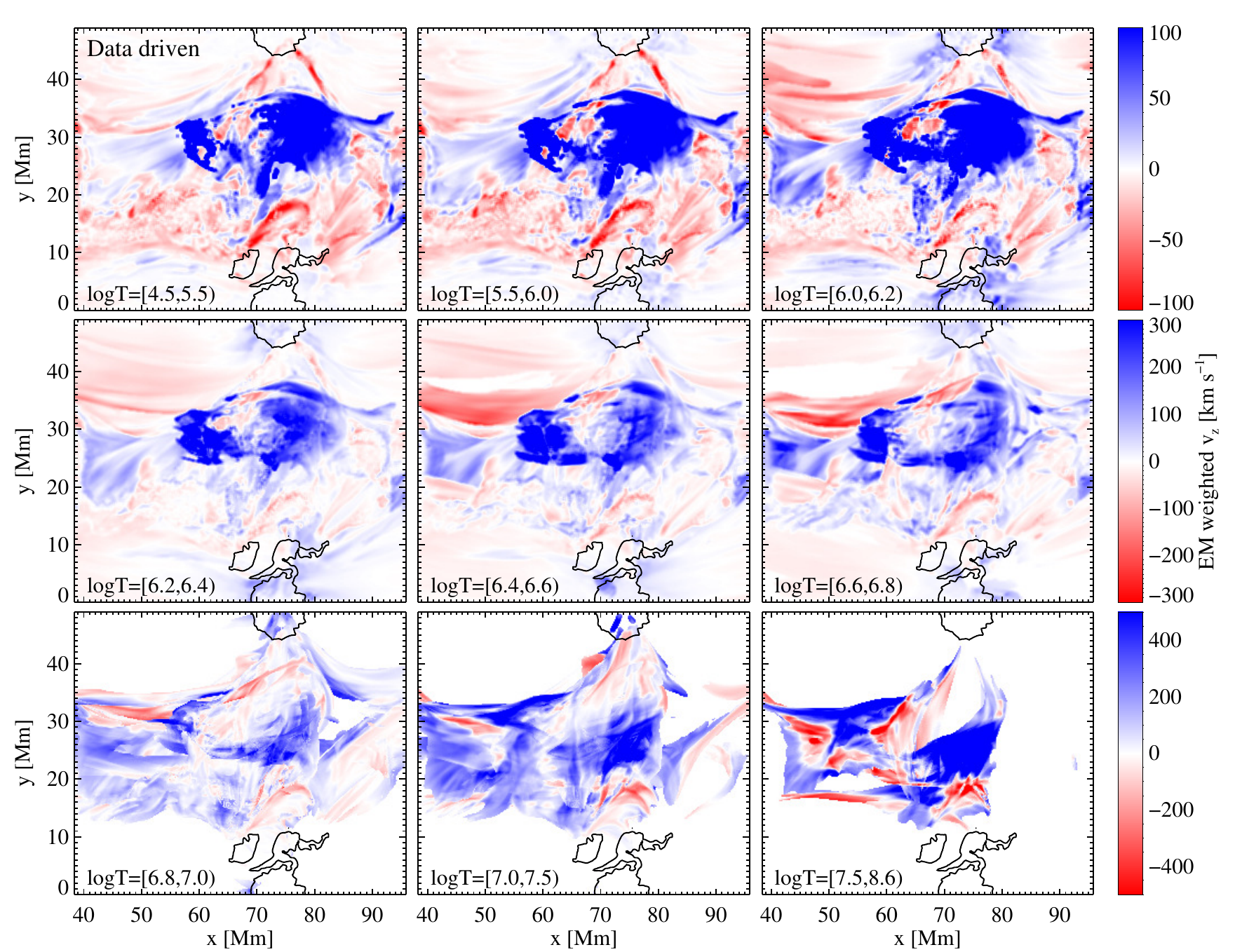}{0.7\textwidth}{}}
\caption{Comparison of the EM weighted $v_{z}$ (i.e., line-of-sight velocity when viewed from the top) of the ground truth (upper panels) flare and data-driven (lower panels) flare run. The black contour lines show $|B_{z}|$ of 1500\,G. The chosen snapshots are identical to those presented in \figref{fig:em} (i.e., 9\,s before the flare peak). The data have been shifted by $L_{y}/2$ in the periodic $y$ direction as done for \figref{fig:vapor}, such that the eruption can be shown in the center of the field-of-view.
\label{fig:vlos_full}}
\end{figure*}

Reproducing plasma motions during a highly dynamic eruption is a nontrivial task. During the flare, both the plasma motion due to the flux rope ejection driven by the Lorentz force and that given rise by evaporation flows driven by the pressure gradient force after the deposition of energy in the lower atmosphere are significant. For comparison with the ground truth flare, we calculate the EM weighted average $v_{z}$ within the 9 temperature ranges shown in \figref{fig:vlos_full}. 

In the low-temperature ($T{{\le}}10^{6}$\,K) panels shown in \figref{fig:vlos_full}, the most evident features are two patches of strong upflows that are close to or greater than 100\,km s$^{-1}$. This is clearly far more than the typical speed in the lower atmosphere and is related to the rise of the cool plasma with the erupted magnetic flux rope. The left patch at $x{\approx}60$\,Mm is cospatial in both simulations, while the right patch is misaligned by slightly more than 10\,Mm. The difference location of the right patch in the ground truth and data-driven runs can also be seen in the side-view EM shown in \figref{fig:em}. A patch of cool plasma is found on the upper-right of the flux rope arch at $x{\approx}$85--90\,Mm in the ground truth flare (\figref{fig:em}(a)). A similar cool plasma patch, or in other words, the top-right corner of the flux rope arch in the data-driven run, appears at $x{\approx}$75--80\,Mm (\figref{fig:em}(b)). This illustrates to some extent how the plasma dynamics are related to detailed structures, and hence reproducing the plasma motions in a point-to-point manner is very challenging.

In the medium-temperature range (middle row in the panel arrays in \figref{fig:vlos_full}), we still the imprint of the plasma carried by the erupted flux rope, i.e., the upflow patches that appear at similar locations as in the low-temperature panels discussed above. The data-driven simulation nicely reproduces the dynamics of coronal loops in the region of $x{{\le}}60$\,Mm. This includes the downward motion in 30\,Mm${{\le}} y {{\le}}$40\,Mm and upflows in 20\,Mm${{\le}} y {{\le}}$30\,Mm. The former are from the long loops connecting the two pre-existing sunspots, as shown in the middle panels of each panel array in \figref{fig:em_full}. The latter is from the foot point area of loops connecting the small positive-polarity magnetic elements and the negative sunspots. Thus, the upflows in this region primarily indicate the evaporation flows along the loops, with a contribution from the projected perpendicular motion of the loops.

In the high-temperature ($T{\ge}10^{7}$\,K) panels, the most prominent feature, which can be seen in the region of 70\,Mm${{\le}} x {{\le}}$80\,Mm and 15\,Mm${{\le}} y {{\le}}$25\,Mm in both the ground truth and data-driven runs, is a patch of strong (${>}500$\,km s$^{-1}$) upflows right next to cusp-shaped strong down flows. They are most clearly visible, particularly for the hottest plasma. This flow pattern manifests the magnetic reconnection that occurs under the erupted flux rope and creates bidirectional outflows. The data-driven run also reproduces the hot upflows in the region of $x{{\le}}60$\,Mm and 20\,Mm${{\le}} y {\le}$30\,Mm, which manifest evaporation flows in the small magnetic element region.

By comparing the EM weighted line-of-sight velocity over the whole temperature, which can be considered a proxy of the Doppler shifts of spectral lines forming in corresponding temperature ranges, we conclude that the data-driven simulation can reproduce in general the key plasma dynamics during the ground truth flare.

\section{Control experiments}\label{sec:a_exp}
\begin{figure*}
\includegraphics{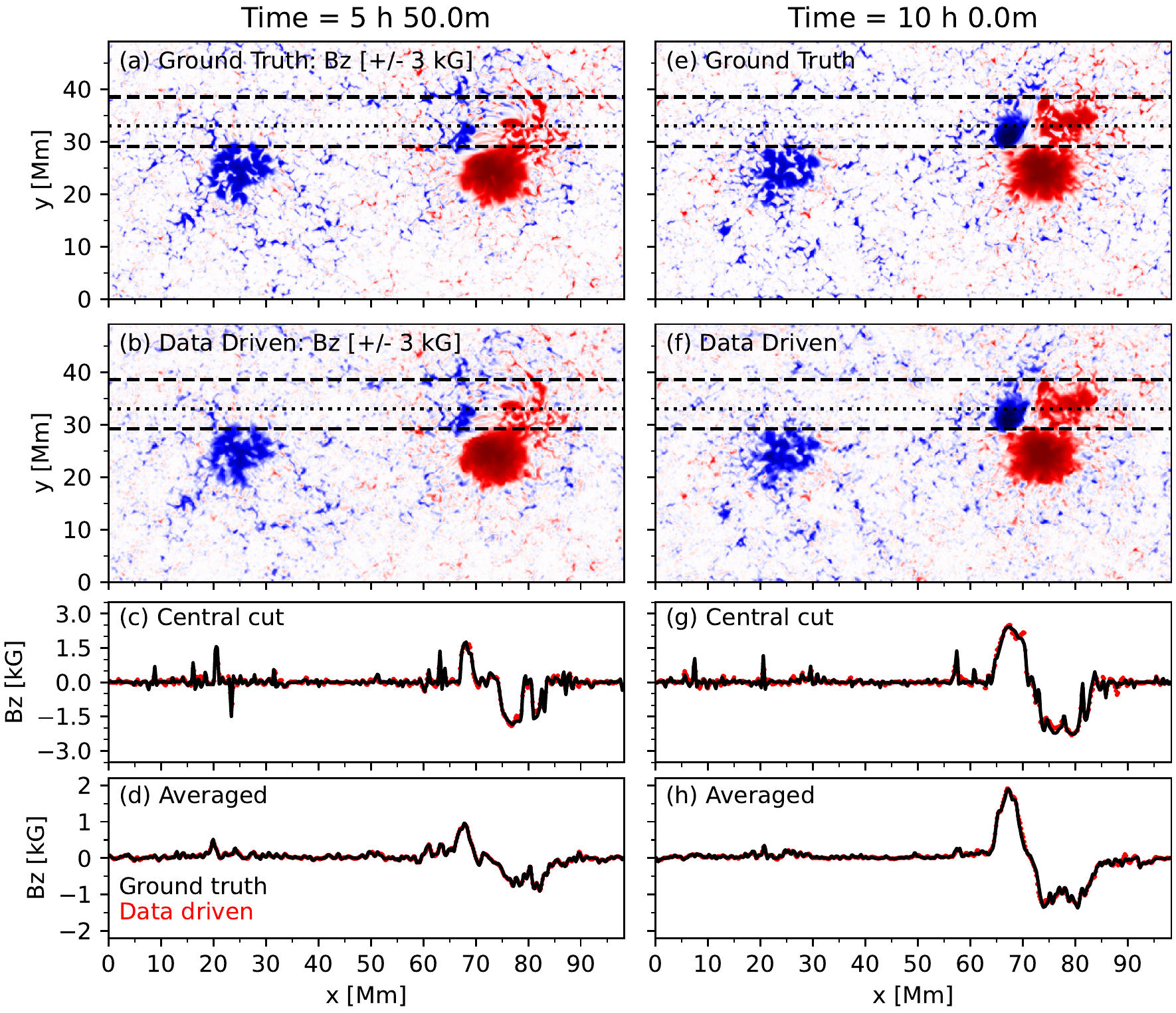}
\caption{Validation of the data-driven boundary in a test with $\Delta z{=}16$\,km. All panels are shown in the same fashion as \figref{fig:compare_bz}.
\label{fig:compare_bz_16}} 
\end{figure*}

\begin{figure*}
\includegraphics[width=18cm]{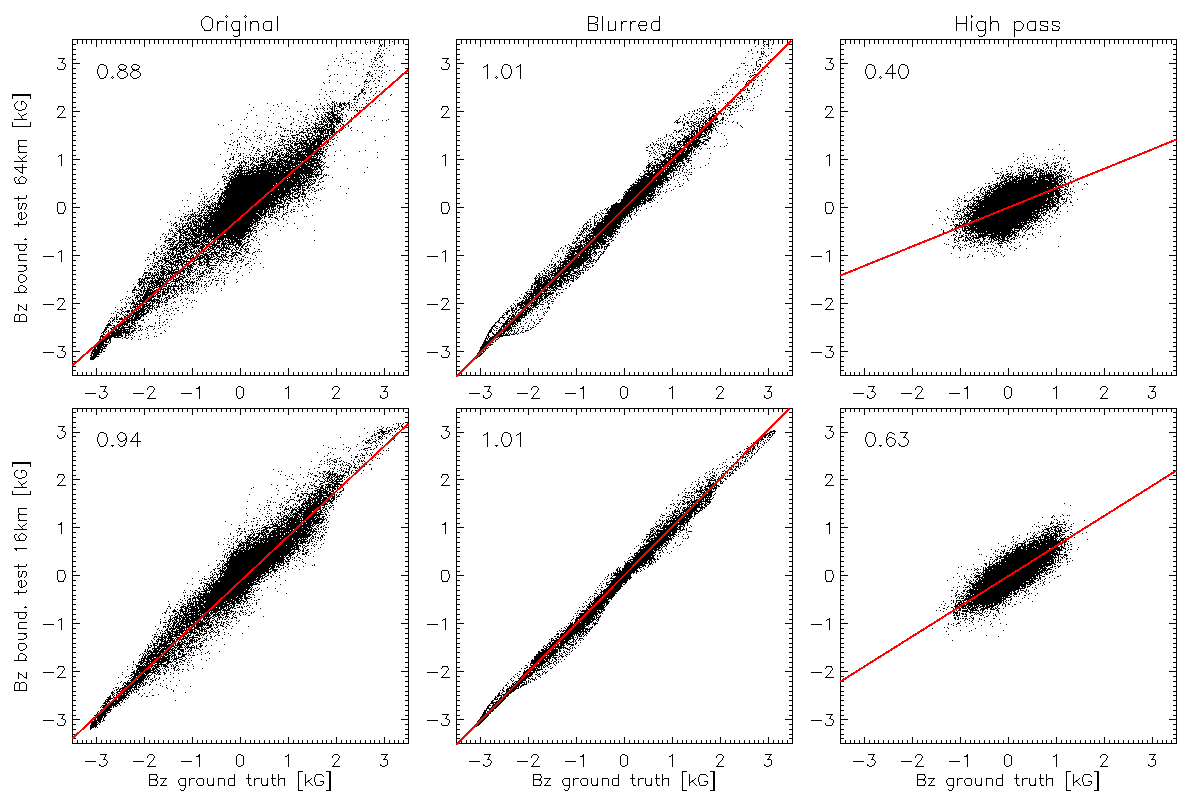}
\caption{Comparison of $B_{z}$ in the boundary tests with $\Delta z{=}64$\,km (upper panels) and $16$\,km (lower panels) with the ground truth $B_{z}$. The snapshot is at $t=10$\,hrs, i.e., the right columns shown in \figsref{fig:compare_bz} and \ref{fig:compare_bz_16}. The red line in each panel shows a linear fitting, with the slope noted in the upper left corner.
\label{fig:compareb}} 
\end{figure*}

\subsection{Boundary test run with finer vertical resolution}\label{sec:a_boundary}
The bottom boundary of the data-driven simulation is defined at the lower face of the first cell in the computational domain. The information of the external data that drives the evolution of the data-driven simulation is placed as a vertical flux through the bottom boundary. This means that the cell-centered physical quantities of the first layer cells are half $\Delta z$ higher than the imposed boundary. This is different from simulations that define both the bottom boundary and magnetic field cospatially at cell interfaces. Another effect, which may be more important, is that the imposed boundary controls the induction equation by a vertical derivative (divergence) of a flux. Meanwhile, the magnetic field may also be influenced by the flux through other faces. Therefore, with a smaller vertical grid spacing, the imposed flux poses a stronger control on the evolution of the (horizontal) magnetic field in the first layer cells and eventually forces the vertical magnetic field to match closer to the ground truth magnetogram.

As shown in \figref{fig:compare_bz}, $B_{z}$ in the first layer cells is already consistent with the ground truth magnetogram under the setup of $\Delta z{=}64$\,km (hereafter the reference test), except for some small-scale magnetic elements and detailed structures at the edge of the sunspots. It is a question whether a finer vertical resolution improves the results as expected. This is validated by a boundary run, as described in \sectref{sec:res_bound}, with a 4 times higher vertical resolution than the reference test. The simulation is set up with the same thin slab domain of $L_{x} \times L_{y} \times L_{z}{=}98.304\times49.152\times2.048$\,Mm$^{3}$ as for the reference test. The vertical domain is resolved by $128$ gridpoints, yielding a much finer $\Delta z{=}16$\,km vertical grid spacing. The other parameters are identical in both runs.

 A comparison between the ground truth and the high resolution data-driven boundary test is presented in \figref{fig:compare_bz_16} in \sectref{sec:a_boundary}. It is clear that the consistency between the data-driven magnetic field and the ground truth magnetogram are greatly improved (they are still 8\,km away in physical height) for both the sunspots and the small-scale magnetic elements through the evolution of 10 hours. In \figref{fig:compareb}, we quantitatively compare $B_{z}$ in the two boundary tests with the ground truth. The blurred data are produced by convolving the original data with a Gaussian filter with $r_{0}{=}5$ ($e$-fold radius${=}r_{0}/\sqrt{2}$). High-pass data are obtained by deducting the blurred data from the original data. Both tests show a good overall correlation with the ground truth. In particular, the slopes of linear fittings for larger structures (the blurred data) in both tests  are close to unity. The improvement of the finer vertical grid spacing on small-scale structures (i.e., the high-pass data) is significant, as reflected by a larger slope and clearly less scattering.

\begin{figure*}
\includegraphics{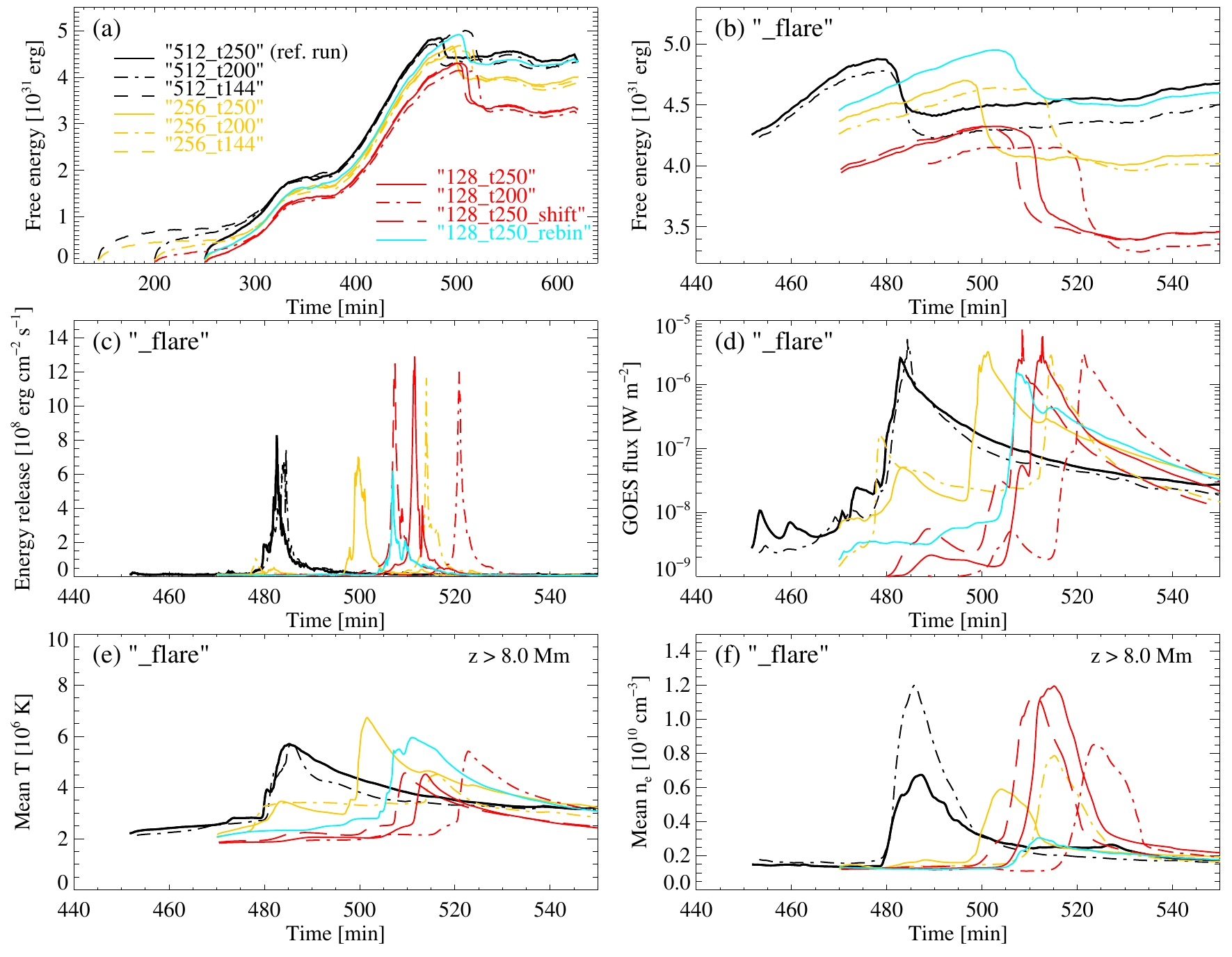}
\caption{General properties of the control experiments with the ``256" and ``128" setups. The correspondence of run names and lines are marked in panels (a). The notation also applies for panels (b) -- (f), which present the flare runs associated with the evolution runs plotted in panel (a).
\label{fig:vstime_appd}}
\end{figure*}

\subsection{Dependence on the horizontal resolution}\label{sec:a_resolution}
We have taken advantage of running the data-driven simulation with the same grid spacing as the ground truth simulation. Observations, from which the data-driven boundary is derived, sample the Sun at a finite spatial resolution that is arguably much lower than the needed value (albeit not yet known) to resolve all details. MHD processes occur on the real Sun under parameters (e.g., the (magnetic) Reynolds number) that cannot currently be achieved by numerical simulations. Therefore, it is worth testing the performance of data-driven simulations with lower resolutions.

For this purpose, we conduct data-driven simulations with 2 times ("256\_t250" run) and 4 times lower resolution ("128\_t250" run), as listed in \tabref{tab:list}. All runs have the same $L_{x}\times L_{y}\times L_{z}=98.304\times49.152\times49.152$\,Mm$^3$ domain. The domain is resolved by $256\times128\times768$ gridpoints in the ``256\_t250" run and $128\times64\times768$ gridpoints in the ``128\_t250" run, resulting in 384\,km and 768\,km grid spacing, respectively. The magnetogram and electric field that are used to drive the simulation are also rebinned such that the low-resolution electric field reproduces the evolution of the low-resolution ground truth magnetogram (flux consesrved). The low-resolution data-driven simulations (evolution runs) start at the same time ($t_{0}{=}250$\,min) as the reference run, and runs starting from different initial times are discussed in \sectref{sec:a_start}.

The low-resolution runs present a similar evolution curve of the free magnetic energy as that in the reference run. However, the amount of free magnetic energy is considerably smaller. We evaluate $\epsilon_{\rm f}$ for the low-resolution runs at $t{=}460$\,min. The ``256" and ``128" runs reproduce 92\% and 82\% of the free magnetic energy in the original high-resolution ground truth, respectively. When compared with the free magnetic energy in the ground truth data that are also rebinned to lower resolutions (reduced to 98\% and 96\%, respectively), $\epsilon_{f}$ would be increased slightly to 93\% and 85\%, respectively. 

The flare does occur in the low resolution runs as well and gives rise to a similar abrupt decay of the free magnetic energy, as shown in \figref{fig:vstime_appd}(a). We conduct the corresponding flare runs, which are started about 30\,min before the flare occurs in the evolution runs. The comparison of the free energy in flare runs in \figref{fig:vstime_appd}(b) indicates that the onset of the eruption is delayed and that the free energy release during the eruption becomes larger when the resolution becomes 2 and 4 times lower. The larger free energy decay in the low-resolution flare runs gives rise to higher flare classes, as shown in \tabref{tab:list}. In particular, the ``128\_t250\_flare" run reaches C5.7, which is higher than the ground truth, although this is achieved by consuming more than twice the free energy. 

A comparison of the magnetic energy release rate is shown in \figref{fig:vstime_appd}(c). The peak of the ``256" flare run reaches a similar strength as the reference flare, and the ``128" flare run has a peak release rate ($1.2\times10^{9}$\,erg s$^{-1}$ cm$^{-2}$) similar to that of the ground truth flare. Although the main flare is delayed in the ``256" flare run, there is a small and noticeable peak at $t{\approx}480$\,min, roughly coinciding with the main flare in the ``512" flare run. 

The synthetic GOES soft X-ray flux curves of the low-resolution flare runs shown in \figref{fig:vstime_appd}(d) have a similar shape as the reference run during the main flare. A hump, which is still lower than B class, is created by the small energy release peak seen in the ``256" flare run. A similar feature can also be found in the GOES flux curve of the ``128" flare run, albeit much weaker, at $t{\approx}490$\,min. However, the low-resolution runs do not reproduce the multiple small peaks (~A class) before the main flare, which are robustly found in all ``512\_" flare runs shown in \figref{fig:vstime}(d).

The mean coronal temperature and density during the flares in the low-resolution runs show very case-dependent evolution, as plotted in \figref{fig:vstime_appd}(e) and (f). The ``256" flare has the highest mean temperature of (close 6.5\,MK) and lowest mean density in all data-driven flare runs, which actually means it creates a weaker CME, as shown in \figref{fig:em-full-256-t250-flare} in \sectref{sec:a_compare_em}. In contrast, the ``128\_t250\_flare" run yields the lowest mean temperature of about 4.5\,MK and the highest mean density, because much more cool plasma is elevated to the corona, as shown \figref{fig:em-full-128-t250-flare}.

We conduct a special experiment  (``128\_t250\_rebin")  that is evolved with the high-resolution mesh ($512\times256\times768$ gridpoints) as the ``512" runs but driven by the low-resolution electric field that is used for the ``128" run. To make the electric field array fit, the $128\times64$ electric field array is expanded to a $512\times256$ array by nearest neighbor sampling, such that one pixel in the low-resolution array becomes $2\times2$ pixels with identical values. As shown in \figref{fig:vstime_appd}(a), the evolution of the free magnetic energy of this run more or less follows the trend of the ``256" run and captures the small transient at $t{\approx}360$\,min, which is not present in other ``128" runs. This means on the one hand, evolving with higher resolution does help this run to produce more free energy and dynamics compared to native ``128" runs; on the other hand, it is obvious that increase the resolution can not compensate the loss of energy contributed by small-scale structures.

The magnetic energy decay during the flare is similar to that of ``512" run, whereas the flare onset time is more similar to the ``128" run. The overall properties of the ``rebin\_flare" run is similar to the weaker flares of ``512" runs. The delay allows the ``rebin" run to accumulate more free energy before the flare occurs. The remaining free energy, as well as its evolution after the flare, appears more close those of the ``512" run. The energy release peak is lower compared to other runs shown in\figref{fig:vstime_appd}(c) and leads to a lower flare class, as well as a weak CME. The flare occurs in this run also bifurcate as a main flare followed by a sub-B class microflare.
     
\subsection{Dependence on the start time $t_{0}$}\label{sec:a_start}
Although the starting time of $t_{0}{=}250$\,min is sufficiently early to capture pre-eruption evolution since the emergence of the parasitic bipole in the photosphere, we also conduct early runs that are started at $t_{0}{=}200$ for all resolutions, including long evolution runs and the corresponding flare runs. These are accompanied by ``512\_t144" and ``256\_t144" runs that are started at $t_{0}={144}$\,min, about 2 hours before flux emergence starts in the photosphere.

As shown in \figref{fig:vstime_appd}(a), the evolution of the free magnetic energy in runs with different start times converges well to the reference run after $t{\approx}300$\,min. Low-resolution early runs converge to the corresponding runs starting at $t_{0}{=}250$\,min. This behavior is consistent with the fact that the growth of the free magnetic energy in the domain is caused by the emerging parasitic bipole. It is also interesting to note that $E_{\rm free }$ in the early runs quickly grows from null to a plateau of about $8\times10^{30}$\,erg, before rising again in response to flux emergence. The plateau corresponds to free energy generated by the initial potential field being stirred by the velocity field. 

The evolution of the free magnetic energy of the ``t200\_flare" runs at all resolutions is similar to their corresponding ``t250\_flare" runs, as shown in \figref{fig:vstime_appd}(b). The flare onset time in the low-resolution early runs is delayed for about 20\,min, while the ``512\_t200" flare does occur at the same time as the reference run. There is no simple trend of how flare energetic may change when the run is started earlier. The profiles of the magnetic energy release of the ``512" and ``128" early runs are similar to the ``t250" runs with slightly lower peaks. The profile of the ``256\_t200\_flare" run becomes more peaky. Nevertheless, the free energy decay does not change significantly, particularly for the low-resolution early runs (+18.5\%. +4.6\%, -7.6\% changes relative to the ``512", ``256", and ``128" flare runs, respectively).

The GOES flux curves of the ``t200\_flare" runs present similar general shapes as their corresponding ``t250\_flare", although the peak fluxes (i.e., flare class) are somewhat different. Scattered behavior is also found in the thermodynamics of the early runs when compared with their corresponding ``t250\_flare" runs. The mean temperature and density profiles indicate that the ``512\_t200\_flare" gives rise to a stronger CME, while still reaching the mean temperature. The low-resolution runs present an anti-correlation between the mean density and temperature, as a stronger CME brings more cool plasma into the corona and leads to a lower mean temperature. A more detailed view of the plasma thermodynamics during the flare peak time by EM is presented in \figsref{fig:em-full-512-t200-flare}, \ref{fig:em-full-256-t200-flare}, and \ref{fig:em-full-128-t200-flare} in \appdref{sec:a_compare_em}.

\subsection{Dependence on run strategy and parameters}\label{sec:a_para}
A parameter we can choose is the maximum flow speed ($v_{\rm max}$) allowed in the data-driven evolution run. The choice of speed limit may also determine the maximum allowed Alf\'ven speed, which adapts dynamically with the maximum sound speed and flow speed as described in \citet{Rempel:2017}. Although the value chosen in the ground truth and reference runs is sufficient for the dynamics of non-eruptive active regions, we explore how the results may change by comparing a series of ``512\_t250\_" runs that have higher $v_{\rm max}$ up to 3000\,km s$^{-1}$. 

Among the tests, two particular cases aim to cover the long-time evolution and flare dynamics in one go. In addition to the ``512\_t250\_v3000" run described in \sectref{sec:method_setup}, we perform the ``512\_t250\_v1500" run with $v_{\rm max}{=}1500$\,km s$^{-1}$ and unlimited temperature. The velocity ceiling is hit for about 60\,s during the peak of the flare. Given that the duration of the energy release peaks shown in \figref{fig:vstime}(c) and \figref{fig:vstime_appd}(c) can last for 20\,min (including the two wings), this run is also considered an evolution-flare run in one. ``512\_t250\_v300" is setup similar to the reference run but with a slightly higher $v_{\rm max}{=}300$\,km s$^{-1}$ because high $v_{\rm max}$ runs suggest that the typical maximum flow speed in non-eruptive stage of the evolution is about 260\,km $s^{-1}$. A flare run is conducted for the ``v300" evolution run to investigate the flare properties.

A comparison of the free magnetic energy and flare properties of the test series with the reference run is shown in \figref{fig:vstime}. A choice of $v_{\rm max}=300$km s$^{-1}$, has almost no impact on the evolution of the free magnetic energy until $t{=}470$\,min. This is because $v_{\max}{=}200$km s$^{-1}$ in the reference run already exceeds the velocity at most of the gridpoints. The general evolution of the two high $v_{\rm max}$ runs are almost identical because it is obvious that the flow speed does not hit the imposed limit for most of the evolution. The most visible difference with the reference run, albeit still small, is found at $t{\approx}360$\,min, when a B class flare (measured in the two high $v_{\rm max}$ runs) occurs and temporally creates high-speed flows. Flare-induced transient high-speed flows can trigger a rise in the maximum Alv\'en speed and consequently an enhanced numerical diffusion \citep{Rempel:2014}. Therefore, it can also be seen in \figref{fig:vstime}(b) that the free energy of the high $v_{\rm max}$ runs is slightly lower (by ${<}2$\%) than that of the low $v_{\rm max}$ runs. 

The high $v_{\rm max}$ runs have similar energy release peaks and give rise to similar flare classes, while the ``v300\_flare" run almost reproduces the profile of the reference run. An obvious difference is that the ``v3000\_flare" run first creates a B2 flare that coincides with the main flare in other cases and a main flare that is delayed by about 5\,min. Interestingly, if we inspect more closely the energy release rate and GOES flux profiles of the other runs shown in \figref{fig:vstime}, a precedent small peak can be discern at $t{\approx}480$\,min, although it is mostly blended with the profile of the main flare. In contrast, the two peaks are separated and give rise to two isolated GOES flux peaks in the``v3000\_flare" run. Similar bifurcation seems to be more common in lower resolution runs, such as the ``256\_t250\_flare" and ``256\_t200\_flare" runs, in which the gap between the two events is more significant. Furthermore, we find that the flare class is better correlated with the highest energy release rate, instead of the free energy decay during the flare. When all 12 flare runs listed in \tabref{tab:list} are considered, this is still evident as a general trend with a moderate scatter. The GOES flux is contributed by hot thermal plasma of $T{>}10^7$\,K as indicated by observations \citep[e.g., ][and references therein]{Aschwanden+al:2015_II,Warmuth+Mann:2016b}. Given the same amount of energy input, a strong impulse, instead of a gentler hump, is favored to heat plasma to higher temperatures.

The mean coronal temperature and density of all runs considered here are very similar before the flare occurs, suggesting that the corona reaches the same energy balance. The ``v300\_flare" run shown in \figref{fig:vstime} generally converges very well to the reference flare. It is also interesting to note that the CME in the ``v300\_flare" run shows the highest similarity when compared with the ground truth simulation for the morphology of emission structures, as shown in \figref{fig:em-full-512-t250-v300-flare}. The CME in the ``v1500\_flare" run (\figref{fig:em-full-512-t250-v1500}) appears very similar to that in the reference run. It leads to a similar peak density but a lower peak temperature. A much weaker CME is produced in the ``v3000\_flare" run. As we can see in \figref{fig:em-full-512-t250-v3000}, the morphology of the eruption is still similar to the ground truth flare over a broad temperature range, but the EMs are considerably lower.

\subsection{Dependence on the initial quiet Sun}
We also explore how the nonhomogeneous structures in the quiet Sun snapshot used as the initial atmosphere may affect the evolution and flare. We conduct an experiment with 4 times binned resolution. The plasma variables in the initial snapshot are shifted by $L_{x}/4$ and $L_{y}/2$ in the $x-$ and $y-$directions, respectively, which means the cube from which the initial condition was adopted is shifted by half of the length in the horizontal directions. Then, the initial active region potential field is computed (as for the ``128\_t250" run) and added to the quiet Sun magnetic field. Properties of the ``128\_t250\_shift" run and the corresponding flare run are plotted in \figref{fig:vstime_appd}. The ``shift" run is highly consistent with the standard setup in all properties, except that the flare occurs about 5\,min earlier. The flare class of the ``shift" run appears slightly higher than that of the standard ``128" run. \figref{fig:vstime_appd}(d) shows that two runs yield highly a similar general GOES flux curve, while the peak flux is achieved by a thin spike on the top of the smoother general profile.

\subsection{The flare onset time}\label{sec:a_time}
\begin{figure*}
\center
\includegraphics{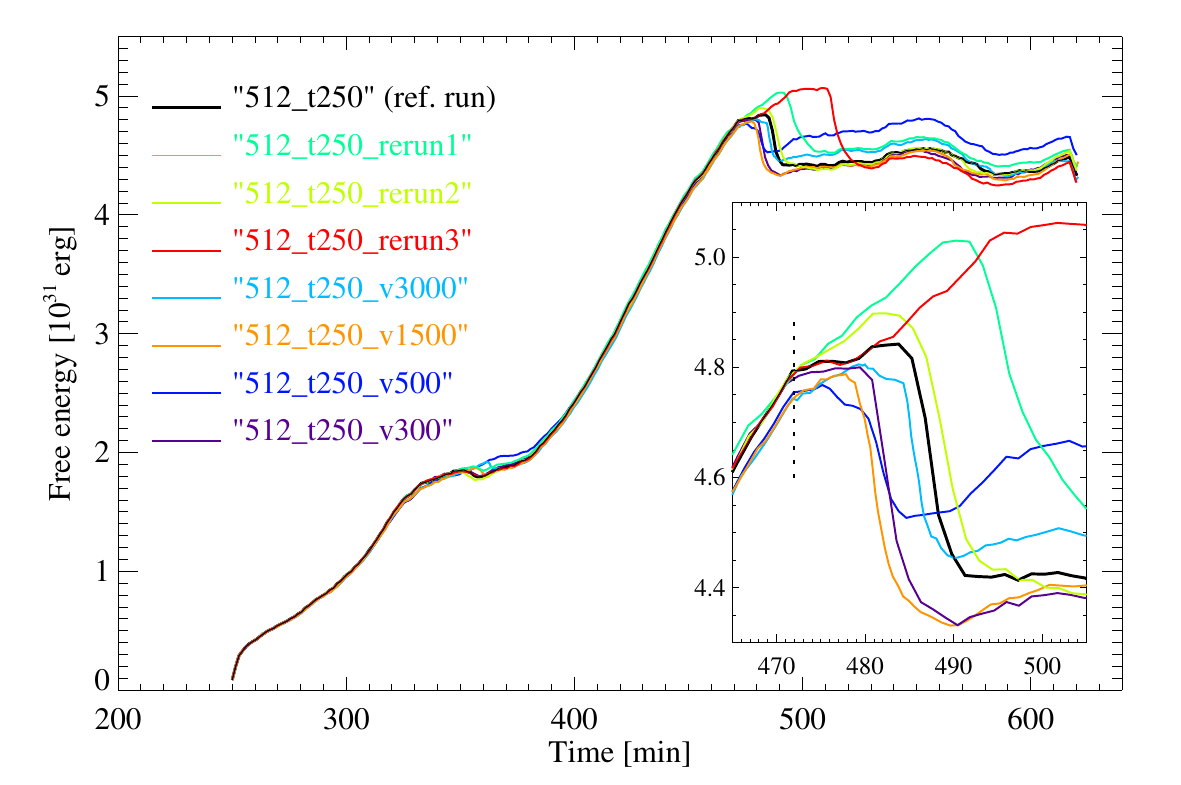}
\caption{The evolution of free magnetic energy in all ``512\_t250" runs. The insert provides a zoomed view of the time period around flare onset. The vertical dotted line in the insert indicates the flare peak time in the ground truth simulation.
\label{fig:flaretime}}
\end{figure*}

\begin{figure*}
\center
\includegraphics{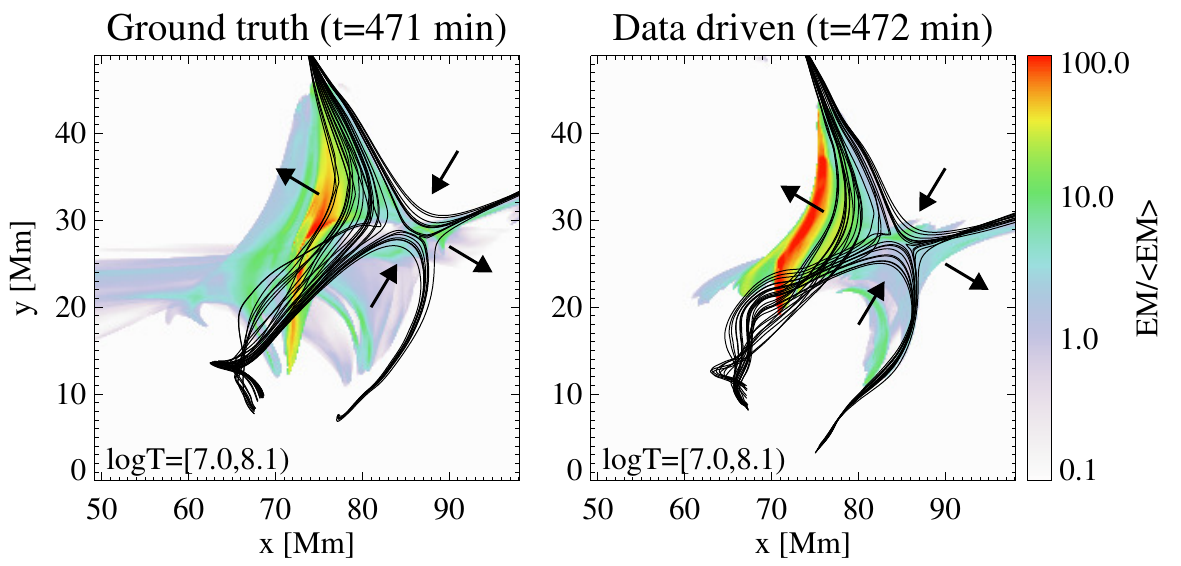}
\caption{EM of plasma hotter than 10\,MK in the ground truth flare and data-driven flare from a vertical view. The EM is normalized by the mean value. The original data have been shifted by $L_{y}/2$ in the $y-$direction. Only the right domain ($x{\geq}L_{x}/2$) is displayed. Black lines are the projection of magnetic field lines traced from seed points randomly distributed in the reconnection region. Arrows indicate an in-/outflow of reconnecting field lines as shown in Supplementary Figure 2 of \citet{Cheung+al:2019}.
\label{fig:flaretime_lines}}
\end{figure*}
The onset/peak time of the flares in data-driven simulations all shows a delay ranging from 8\, to 30\,min compared with the ground truth flare. Although this delay is still much smaller than the 4--6 hour evolution time before the flare, we conduct more experiments including one more evolution run with $v_{\rm max}{=}$500\,km s$^{-1}$ and rerunning the reference run three times to investigate the performance of timing prediction.

The evolution of the free magnetic energy of all ``512\_t250" runs is compared in \figref{fig:flaretime}. All runs consistently follow the same evolution curve with errors less than 2\%. In contrast, the behavior of eruptions, including the main flare and the early small flare at  $t{\approx}360$\,min, are case dependent. Even for reruns of identical setup of the reference run (``\_rerun1", ``\_rerun2", and ``\_rerun3" in \figref{fig:flaretime}), the evolution of the free energy during the non-eruptive stage is almost precisely reproduced; however, the flare onset time can be considerably different, which implies that the trigger of the reconnection has a stochastic component.

The most interesting finding in \figref{fig:flaretime} is a decrease in the growth rate (i.e., the slope of the curve) of the free energy shortly after $t{=}470$\,min. The change in slope is more clearly visible in the insert panel in \figref{fig:flaretime}, as indicated by the vertical dotted after which the consistent curves of all runs begin to diversify. Furthermore, the dashed line actually marks $t_{peak}=472$\,min of the ground truth run. The decrease in the slope suggests that an enhanced release of the magnetic energy occurs in all data-driven runs precisely at the moment when the ground truth flare should be triggered. 

We inspect the EM in the flare runs (as only the flare setup captures the high temperature plasma) at this moment, when the ground truth flare is in the very initial phase of eruption and the data-driven flare run shows a sub-B class precursor. In \figref{fig:flaretime_lines}, we present the vertical view of the EM for $T{\geq}10^7$\,K, which shows hot plasma created by the reconnection, and magnetic field lines traced from seed points that are randomly distributed around the reconnection region. The domain has been shifted by $L_{y}/2$ in $y-$direction, such that the eruption is located in the center of the field of view. In the ground truth flare, reconnection occurs between the field lines anchored in the parasitic bipole and those connecting the pre-existing bipole though periodic boundary (Subdomain 1 and Subdomain 3 as named in Supplementary Figure 2 of \citet{Cheung+al:2019}) as indicated by the arrows in \figref{fig:flaretime_lines}. This is the very first phase of the eruption, as shown in Supplementary Figure 3 and more specifically in Supplementary Figure 4(a) of \citet{Cheung+al:2019}. The exact same process is reproduced in the data-driven flare at most the same time, as illustrated by the highly similar magnetic topology and emission features shown in \figref{fig:flaretime_lines}.

In the ground truth flare, the eruption immediately develops to the next stages (i.e., the reconnection discussed earlier \sectref{sec:res_bfield}), which leads the flare into the peak. In contrast, the further transition is slightly delayed in the data-driven flare, although it again follows the reconnetion processes similar to those in the ground truth flare and develops to the flare peak. The transition can also be detected in \figref{fig:flaretime_lines}: some hot emission has become visible in the ground truth flare, as a branch extending to $x{\leq}65$\,Mm, whereas a corresponding feature is basically invisible/insignificant in the data-driven simulation at this moment. 

Nevertheless, we conclude that at least a precursor-like event could be predicted at almost the correct onset time of the flare and that the delay in flare peak time is caused by the delay in linking the multiple stages of the reconnection between different flux systems. A similar behavior is also discernible in the lower resolution runs shown in \figref{fig:vstime_appd}(a). We also note that the onset time of the ground truth flare may also be different (delayed by 5\,min) if the simulation is rerun at a higher resolution. We suspect that the onset of a major flare (i.e., an abrupt and significant decay of free magnetic energy) may depend on the nonlinear development of instability in, for example, the current sheet built up between the emerging parasitic bipole and the pre-existing magnetic field. A nonlinear system may not be point-to-point identical after a substantially long evolution time, and hence, the time and place at which the instability is triggered and the subsequent development may change. 

\section{Summary and Discussion}\label{sec:sum} 
We present the method of data-driven radiative MHD simulation with the MURaM code. We use a time series of electric fields derived from the photospheric data of the ground truth simulation as the bottom boundary of a data-driven simulation. Even though the reference run presents an optimistic scenario, where we know exactly how the ground truth is generated, we construct the data-driven simulation from nothing but information that can be derived from observations. When the data-driven run is set up with identical resolution and similar parameters, it reliably reproduces many of the general and detailed properties of the ground truth, including the growth and decay of the domain-integrated free magnetic energy, emission features over a wide range of temperatures, and dynamics of coronal loops involved in the flare. The predicted flare onset time is about 8\,min delayed from the ground truth, which is sufficiently small compared with the 4-hour evolution in advance of the flare, and a flare precursor is found at almost the accurate time of the ground truth flare. Therefore, we conclude that the data-driven simulation can successfully reproduce the ground truth with high fidelity. The robustness of the method and results are validated by a series of control experiments. The control experiments demonstrate an agreement in the general course of the development of free magnetic energy and the eruption. In particular, we have presented that runs using a significantly lower resolution can still capture many of the key properties of the ground truth simulation over a long time period, which endorses the usefulness of the method presented in this paper on real observation data and reproducing observed solar eruptions.

\subsection{Assessing the performance of data-driven simulations}
Assessment of the performance of data-driven simulations via a ground truth has also been elaborated in several studies. \citet{Leake+al:2017} examined how accurate two-dimensional simulations extending from the photosphere to the corona reproduce the coronal free magnetic energy, when they are driven by all variables of ground truth simulations that capture the emergence of twisted magnetic flux rope from the convective region into the corona. They found that low errors can be achieved for cases where the time scale of flux emergence is substantially longer than the driving cadence. \citet{Toriumi+al:2020} compared how well different data-driven methods/codes reproduce the coronal magnetic field properties of a ground truth simulation of flux emergence from the convection to the corona. The results scatter considerably. This test indicated the non-negligible impact from assumptions made in different models on thermodynamic quantities (or parameters that determine how the system responds to magnetic driving force).  \citet{Jiang+Toriumi:2020} noted that driving simulations with data from a higher layer, where the magnetic field is closer to a forcefree state, helps to improve the results. Recently, \citet{Inoue+al:2022} presented a newly developed code for data-driven MHD simulation under zero-$\beta$ approximation and validated the method with a ground truth simulation of flux rope formation through twisting sunspots. The data-driven simulation captures the evolution of the flux rope from formation to eruption caused by sunspot rotation and flux cancellation.

The simulations presented in this study described the processes of flux emergence and a C class flare accompany by a CME, which combine the scope of the studies above. In the best case (only sufficient resolution is required) scenario, the free magnetic is close to the ground truth with a goodness similar to the models of \citet{Leake+al:2017}. The magnetic field in the data-driven simulations also matches the ground truth in most cases. This benefits from using the same numerical code and that the data-driven simulations are constructed to represent a realistic density and temperature stratification as the ground truth does. Last but not least, an important aspect of validation, which was often omitted in previous studies, is how well a lower resolution run reproduces a high-resolution ground truth, because all numerical simulations done thus far are arguably of insufficient resolution  (in terms of, e.g., Reynolds number) to reproduce the real Sun. In our study, the lowest resolution tested is of 768\,km grid spacing, which is comparable to commonly used observational data, and these tests can still capture the flare.

\subsection{Comparing data-driven simulations with observations}
Data-driven simulations have been quite successful in reproducing dynamic features in observed active regions and eruptions. Because many previous simulations have ignored or simplified the treatment of plasma thermodynamics, the comparison with observations relies on relating emission features to magnetic features and derivatives in simulations under certain assumptions. A common practice, which seems to be an intuitive option under the frozen-in condition, is following magnetic field lines that are cospatial (when projected) with the emission features in images. However, plasma hosted in a magnetic structure (e.g., a flux rope) can be multi-thermal as found in numerical models and observations \citep[e.g., ][ and references therein]{Cheng+al:2014,Xia+Keppens:2016,WangCan+al:2022,LiLeping+al:2022}. Although general agreement is quite often found between selected magnetic field lines and observed emission features on large scales, the gap between the simulated and real magnetic fields might be larger than expected because the emission structures (usually seen in a particular passband) reveal only limited details of magnetic topology.

Proxies derived from the magnetic field under further physical assumptions are also used to link features in simulations and observations in a more sensible fashion. For example, quasi-separatrix layers are suggested to correspond to strong current sheets where reconnection may be triggered, and they are thus connected with flare ribbons \citep{Aulanier+al:2005}.\citet{JiangChaowei+al:2016nc,Price+al:2019} showed that the imprint of the quasi-separatrix layers of the simulated coronal magnetic field matches the brightening in AIA 304 and 94\,\AA~ images. \citet{YanXiaoli+al:2022} showed that horizontal maps of quasi-separatrix layers in a data-driven simulation of a confined flare are cospatial with the flare ribbons observed in the H$_\alpha$ line. These studies provide evidence that data-driven simulations capture the key magnetic structures relevant to energy release during these flares.  \citet{Cheung+DeRosa:2012} derived an emission proxy that is related to the mean of current squared along each magnetic field line. With a great saving in computational expense, this approach mimics the effect of Ohmic heating in coronal loops that is treated in a self-consistent manner in realistic coronal simulations \citep{Warnecke+Peter:2019}. When this quantity is integrated along the line-of-sight as optically thin emission in the corona, the resulting synthetic images display numerous loop-like structure and appear highly similar to EUV observations in MK temperature. \citep{JiangChaowei+al:2016} employed a similar method and found that bright features in such a emission proxy appears in line with bright loops observed in AIA 94\,\AA.

The simulations presented in this study account for the energy balance, as done in realistic coronal simulations, in an eruptive model. Therefore, it allows for comparing self-consistently generated observables with actual observations. For example, flare ribbons are directly identified as footpoints of magnetic field lines connecting to the reconnection region and are formed physically as energy is deposited in the lower solar atmosphere. As in this paper, data-driven simulations can be compared with ground truth by synthetic EUV images as well as EMs over a broad temperature range. When applied to real events, this can also help to diagnose the thermodynamic properties of erupted plasma and their relation to magnetic structure.

Reproducing the onset time, particularly after a substantial quasi-static evolution, is an important goal of data-driven simulation. This ability not only suggests that the triggering of the eruption is perhaps resolved to a considerable extent,  but also elevates the reliability of predictions. Although making a precise prediction of the onset time of a real eruption seems difficult for now, some studies have achieved promising results. \citep{JiangChaowei+al:2016nc} presented a simulation covering an evolution of 60 unit-time, in which the eruption is delayed by 2 unit-time compared with observed eruption. A unit-time corresponds to 90\,s in their simulation and is equivalent to 1\,hr solar time, given that the evolution of the system has been artificially accelerated by 40 times.  \citep{GuoYang+al:2019} initialed a data-driven simulation at 1 hour before a flux rope eruption was observed. The eruption in the simulation occurs 20\,min earlier than the actual event. \citep{Kaneko+al:2021} evolved a data-driven simulation for about 300\,min solar time, during which two eruptions of flux ropes are captured. The two eruptions in the data-driven simulation occur about 40\,min earlier/later than the observed events.

The flares that occur in the simulations presented in this study show a delay in the range of 8 to 40\,min after a data-driven evolution in the range of 220 to 330\,min. We also note that, for comparison of the timing, we measure the peak time in the GOES flux curve, as this can be precisely determined, instead of the so-called onset time. Moreover, we find a precursor that occurs at almost the correct onset time in the same place where the ground truth flare is triggered. It may be arguable that producing a real event is more challenging than reproducing a simulated ground truth. The performance of the reference run is similar to the model of \citep{JiangChaowei+al:2016nc}, in terms of the ratio of the time difference (of the main flare) to the evolution duration.  This suggests these models can capture the correct structure that triggered the eruption, as also noted by \citep{JiangChaowei+al:2016nc}.

\subsection{Challenges on application on real data}
Deriving the electric field from observations is an integral part of constructing data-driven MHD models and a non-trivial task. Some extensively used methods infer the velocity field by either directly tracking the apparent motion of line-of-sight magnetic field features \citep[e.g., ][]{November+Simon:1988,Fisher+Welsch:2008} or by incorporating the tracking with solving the induction equation \citep[e.g,][]{Kusano+al:2002,Welsch+al:2004,Longcope:2004,Schuck:2006,Schuck:2008}. An alternative line of methods directly inverts the electric fields that are consistent with the temporal evolution of the magnetic field \citep{Fisher+al:2010,Kazachenko+al:2014,Fisher+al:2020}. In addition to reproducing the magnetic field, additional challenges arise from capturing the energy and helicity fluxes that energized the corona\citep{Welsch+al:2007,Kazachenko+al:2014}. In practice, simulations driven by electric fields derived from observations have indicated that a non-inductive component in the electric field is important for the formation of flux ropes, as discussed in \citet{Lumme+al:2022}. In this study, we follow the method used by \citet{Cheung+DeRosa:2012} when deriving the electric field. We take the advantage of knowing comprehensive information of the velocity and magnetic fields when dealing with the freedom in the horizontal electric field $E_{h}$ when constrained only by the vertical magnetic field, as discussed in \citet{Cheung+DeRosa:2012}, by letting $\nabla_h \cdot E_{h} = \nabla \cdot E^{vb}_{h}$.

Nearly all electric field inversion methods assume that the magnetic and velocity fields are observed on a constant height (or radius) surface, which also facilitates implementation of the electric fields in data-driven simulations. An uncertainty is introduced by the fact that observations measure the magnetic field on a surface of constant optical depth. It is also arguable that a ever higher cadence is desirable to capture fast changes in the magnetic field. However, actual observation is made by HMI \citep{HMI}, for example, every 45 (135)\,s as a high cadence or 720\,s for data of lower noise \citep{Hoeksema+al:2014}.  By driving the simulations presented in this paper with the electric field at 60\,s cadence, instead of the ultimately high cadence (every 50 iterations ${\approx}$0.5--2\,s) available in the ground truth data, we have intended to mimic the cadence of observation to some extent. Further investigation on the possible uncertainty when using real observational data, as discussed here, is within the scope of an ongoing project that will more comprehensively assess the performance of data-driven/constrained modeling of solar eruptions.
 
\begin{acknowledgments}
We thank the referee for helpful suggestions and Dr. Yuhong Fan for comments that improve the clarity of the paper. F.C. is supported by the National Key R\&D Program of China under grant 2021YFA1600504 and the Program for Innovative Talents and Entrepreneurs in Jiangsu. This material is based upon work supported by the National Center for Atmospheric Research, which is a major facility sponsored by the National Science Foundation under Cooperative Agreement No. 1852977. This work benefits from discussions during the ISEE/CICR Workshop on ``Data-Driven Models of the Solar Progenitors of Space Weather and Space Climate" supported by Nagoya University and ISSI workgroup ``Data-driven 3D Modeling of Evolving and Eruptive Solar Active Region Coronae". We would like to acknowledge high-performance computing support from Cheyenne (doi:10.5065/D6RX99HX) provided by NCAR’s Computational and Information Systems Laboratory, sponsored by the National Science Foundation. F.C. was supported by the George Ellery Hale postdoctoral fellowship at the University of Colorado Boulder (2017-2019), when this work was initiated. The visualization shown in \figref{fig:vapor} is created by VAPOR \citep{vapor}.
\end{acknowledgments}

\bibliography{reference}
\bibliographystyle{aasjournal}

\begin{appendix}

\section{Comparison of the EM near the flare peak}\label{sec:a_compare_em}
\begin{figure*}
\gridline{\fig{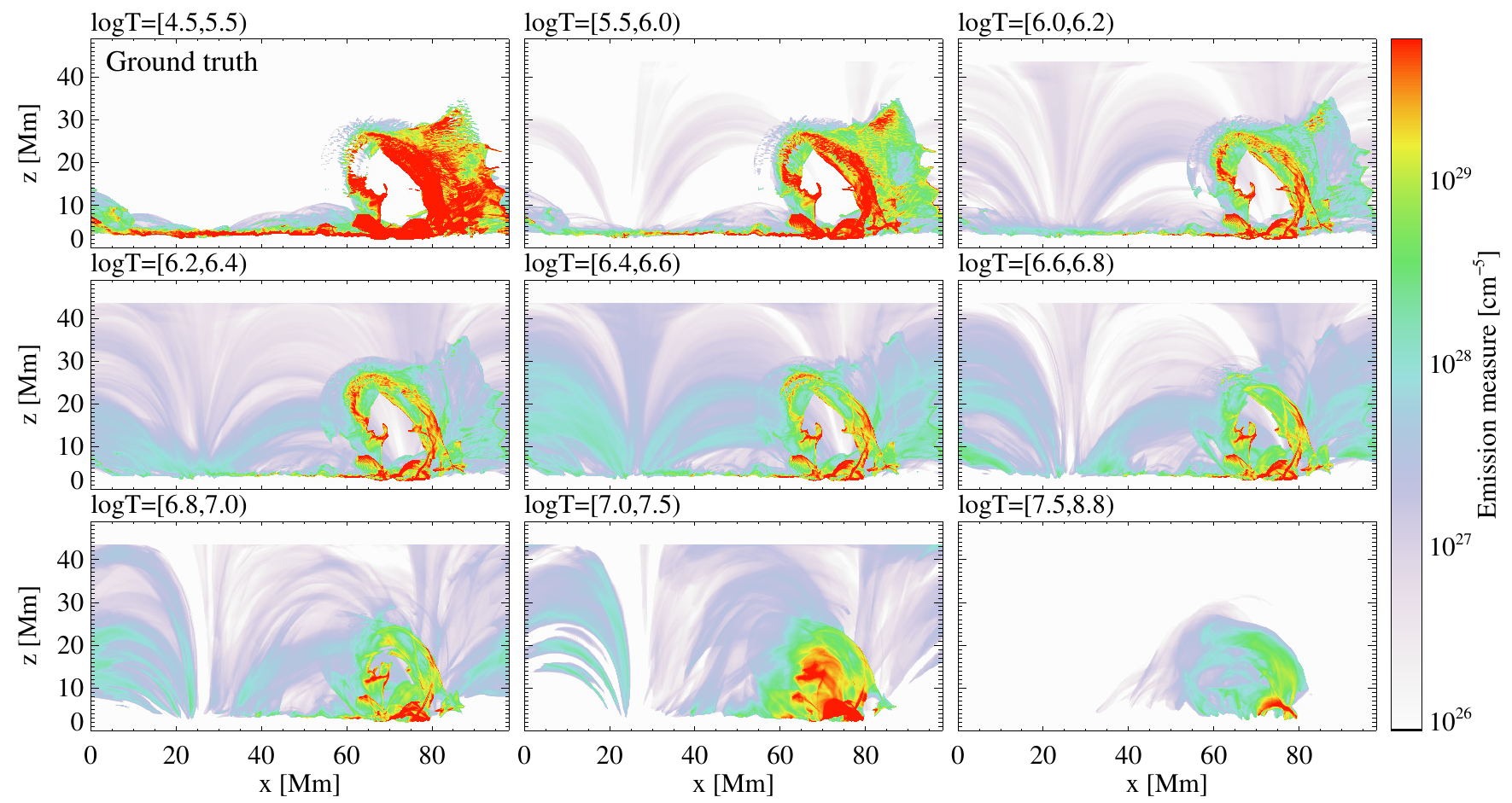}{\textwidth}{}}
\gridline{\fig{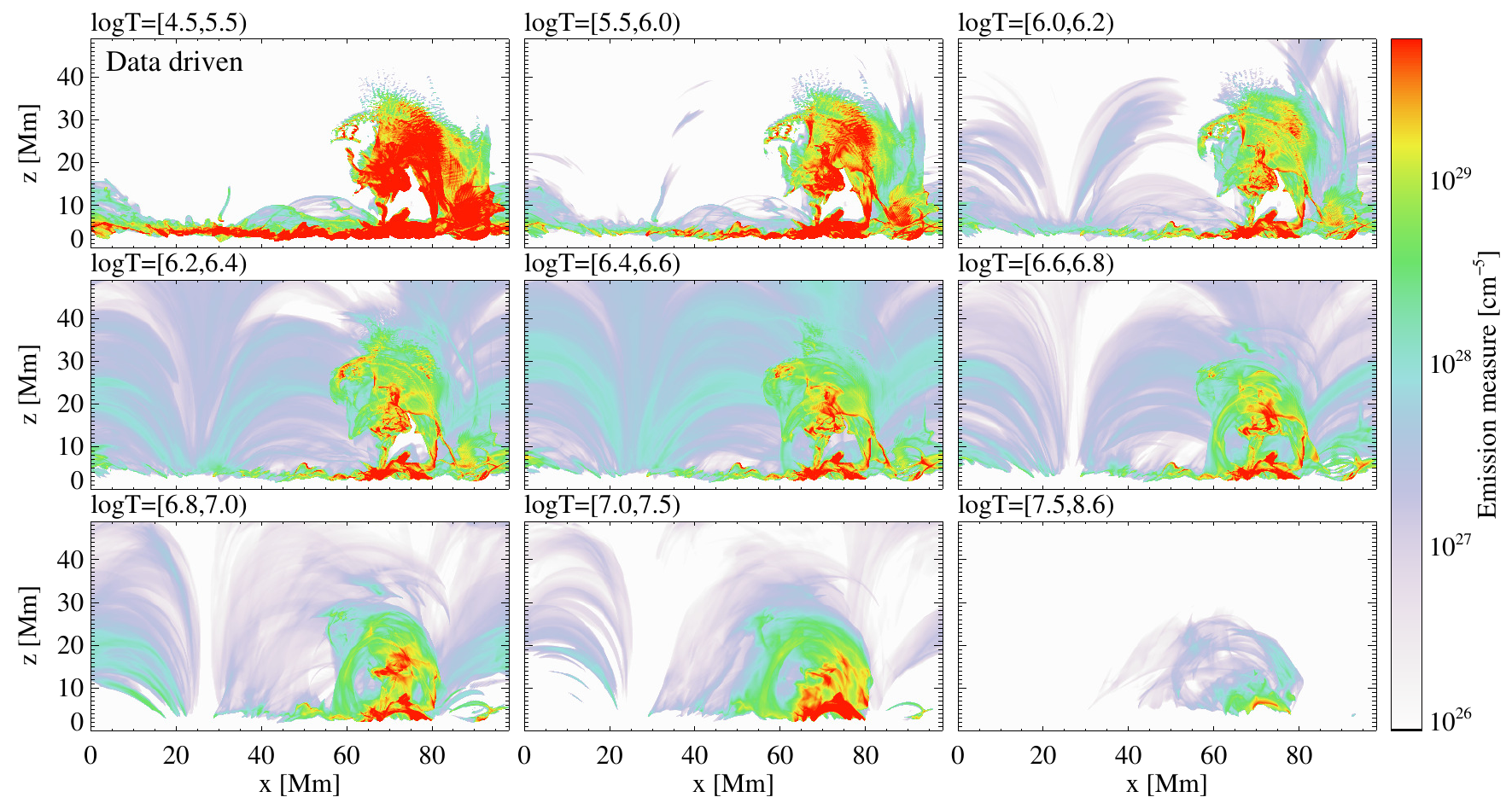}{\textwidth}{}}
\caption{Comparison of the EMs (viewed along the $y$-axis) of the ground truth flare (upper panels) and data-driven (lower panels) flare run. The chosen snapshots are identical to those presented in \figref{fig:em} (i.e., 9\,s before the flare peak).
\label{fig:em_full}}
\end{figure*}
We present a comparison of the EMs of the ground truth flare and data-driven flare run over the full temperature range in \figref{fig:em_full}. The MURaM code stores EM within an interval of $\Delta\log T=0.1$ from $\log T=4.5$ to the highest temperature in the domain. We rebin the EM into 9 temperature ranges, as shown in \figref{fig:em_full}. The ranges are identical for both runs, except the hottest bin where the ground truth flare reaches a higher temperature than the data-driven flare run.

The EM in the lower temperature ranges (top rows in both panel arrays) illustrate the arches of the cool plasma carried by the erupted flux rope, as we have presented in \figref{fig:em}. The overall arch-like shape is similar in both runs, although the structure in the ground truth simulation is much smoother than that seen in the data-driven simulation. The comparison in detail also indicates some differences, such as the top-right corner ($x{\approx}85$\,Mm and $z{\approx}30$\,Mm) of the arch, where the ground truth has a patch of enhanced EM, and the lower-left corner ($x{\approx}70$\,Mm and $z{\approx}15$\,Mm), where the data-driven flare shows more EM.

The EM at medium temperatures (several MK) shows some imprint of the structures seen at lower temperatures, as some of the cool plasma rising with the flux rope is heated to MK temperatures. Moreover, it is very interesting to note that many of the coronal loops in these temperature ranges appear very similar in not only the location but also the exact temperature ranges they show up. This result suggests that the data-driven simulation produces the connectivity of the large-scale magnetic field between sunspots, which is satisfying but not surprising and yields approximately the correct amount of coronal heating that determines the density and temperature of these coronal loops.

The EM at high temperatures generally presents the same results as shown in \figref{fig:em} but in view of small temperature bins. Both flare runs yield hot plasma in loops that are involved in magnetic reconnection, particularly the short post-flare loops above the parasitic bipole. It also helps to quantify that the data-driven flare produces less hot plasma than the ground truth flare.

\begin{figure*}
\includegraphics{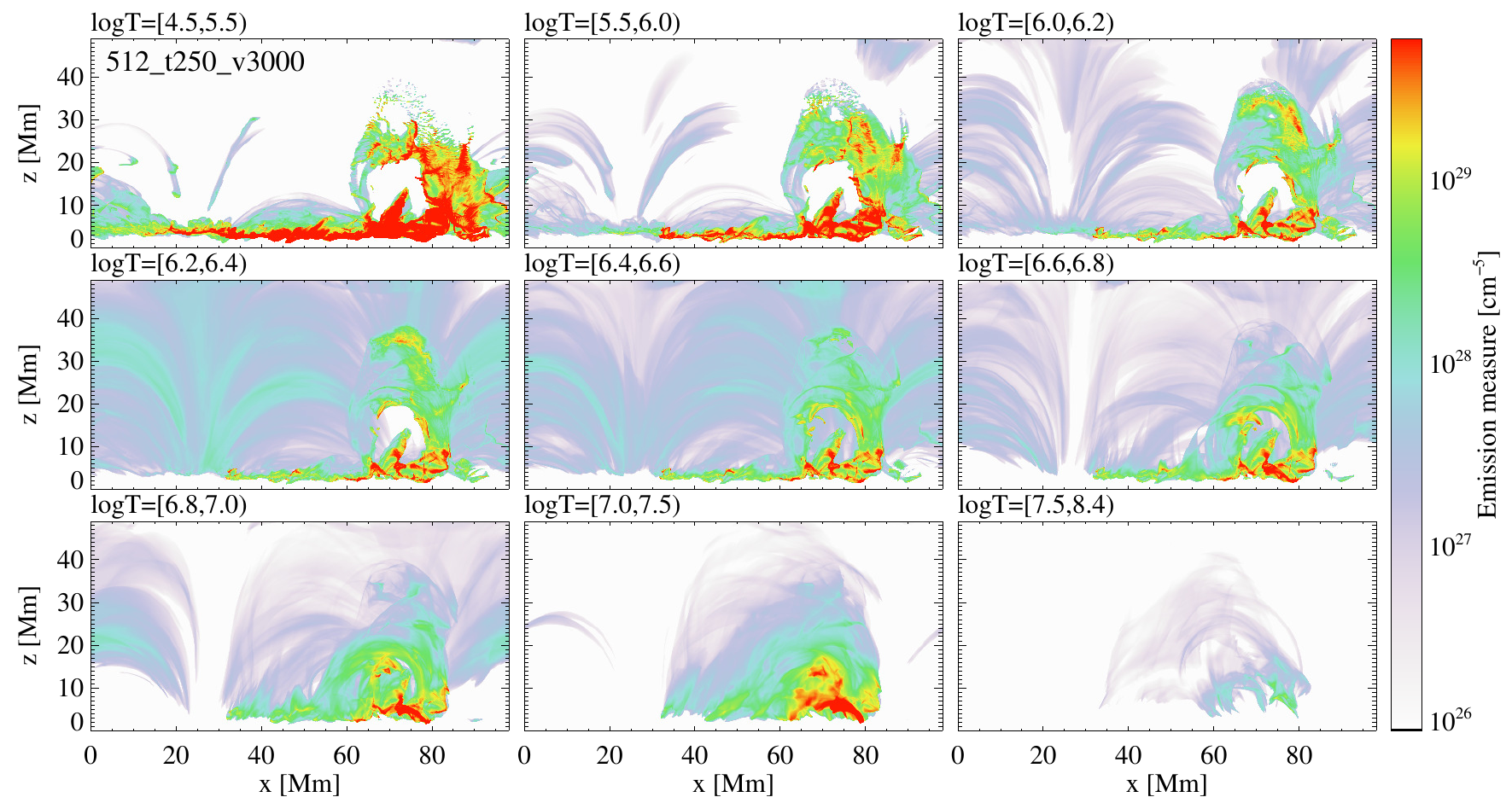}
\caption{EMs (viewed along the $y$-axis) of the ``512\_t250\_v3000" run. The snapshot shown is at 1\,min before the time of the peak GOES flux of this run.
\label{fig:em-full-512-t250-v3000}} 
\end{figure*}

\begin{figure*}
\includegraphics{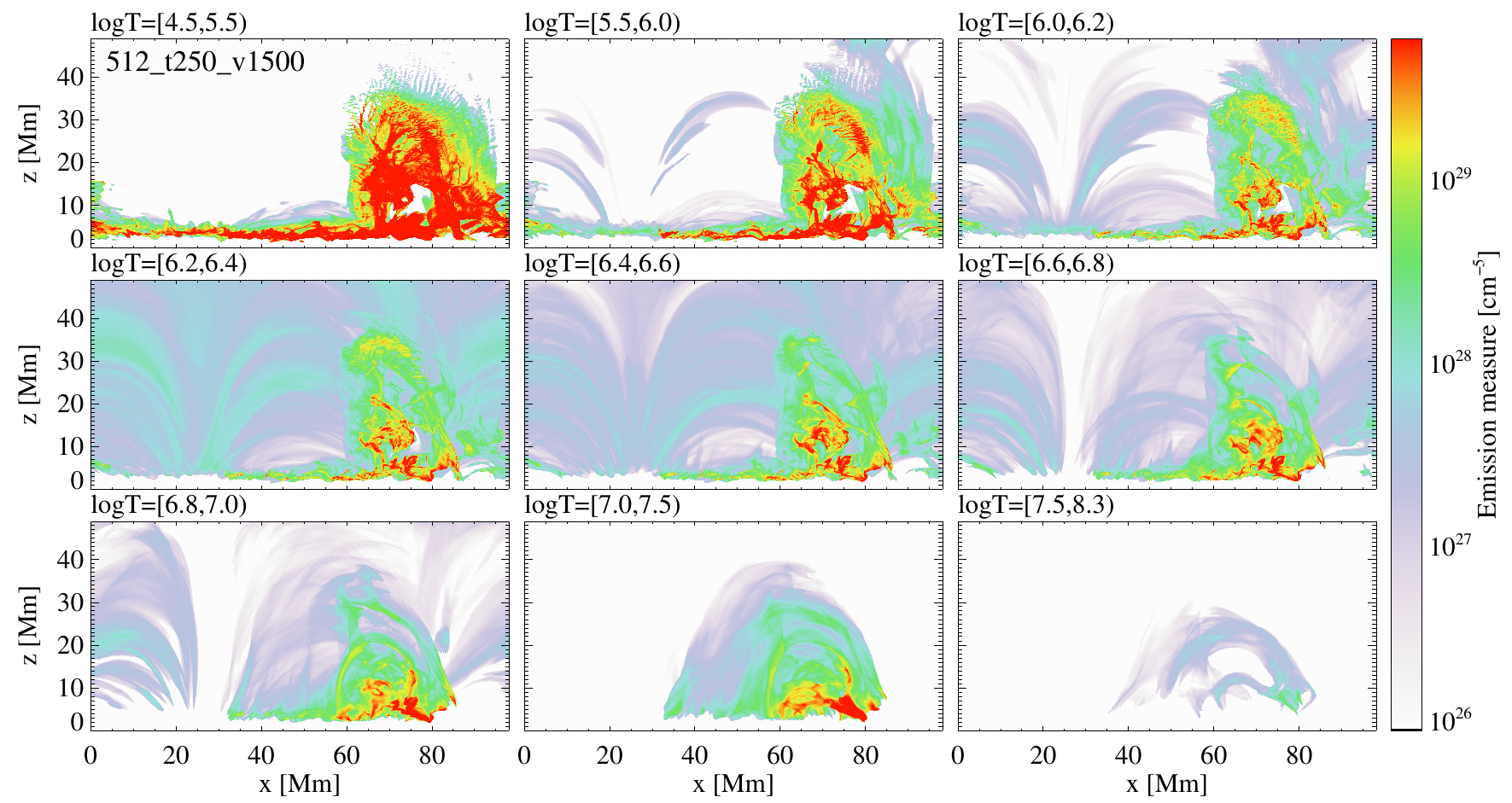}
\caption{EMs (viewed along the $y$-axis) of the ``512\_t250\_v1500" run. The snapshot shown is at 9\,s before the time of the peak GOES flux of this run.
\label{fig:em-full-512-t250-v1500}} 
\end{figure*}

\begin{figure*}
\includegraphics{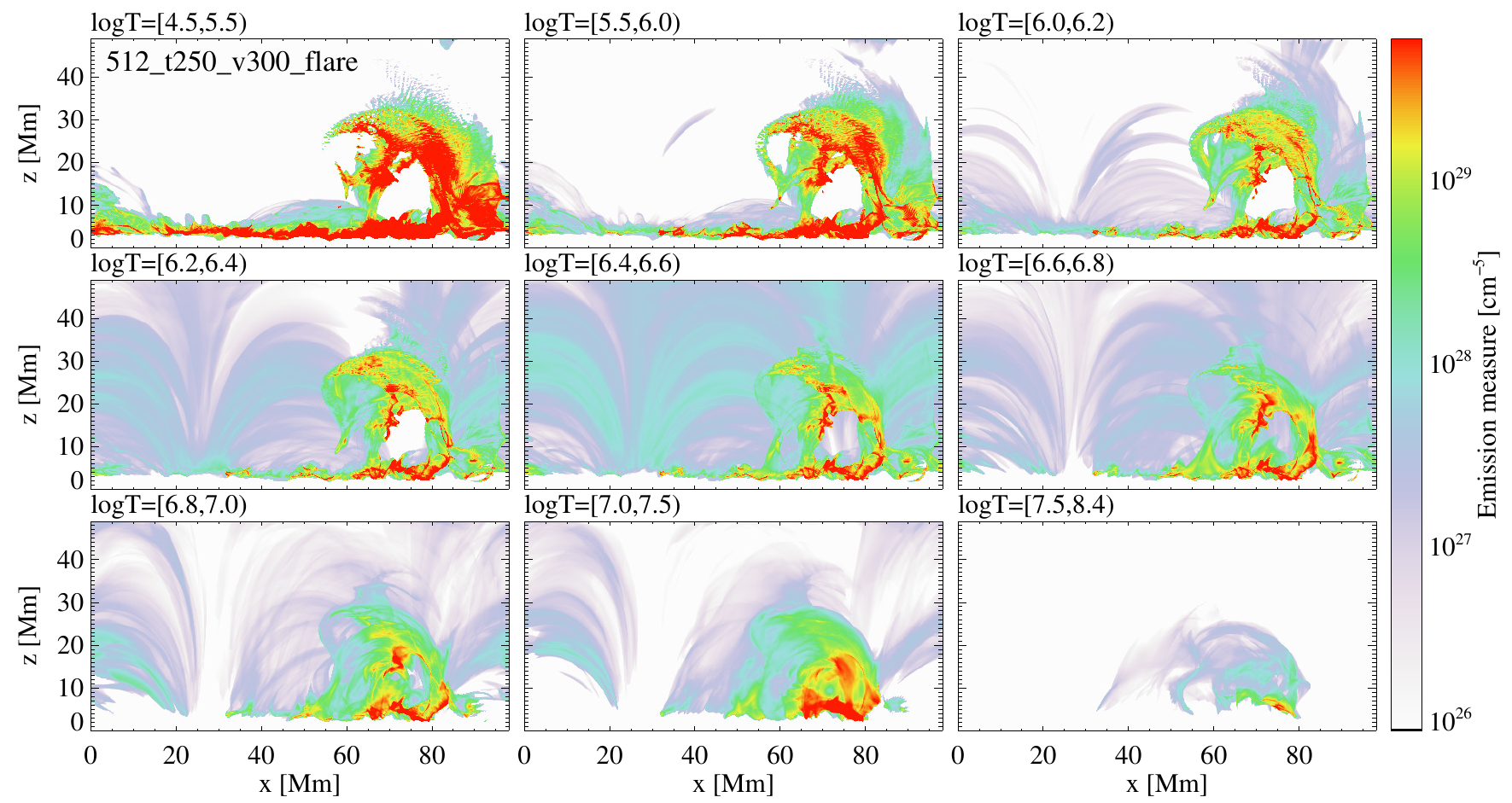}
\caption{EMs (viewed along the $y$-axis) of the ``512\_t250\_v300\_flare" run. The snapshot shown is at 9\,s before the time of the peak GOES flux of this run.
\label{fig:em-full-512-t250-v300-flare}} 
\end{figure*}

\begin{figure*}
\includegraphics{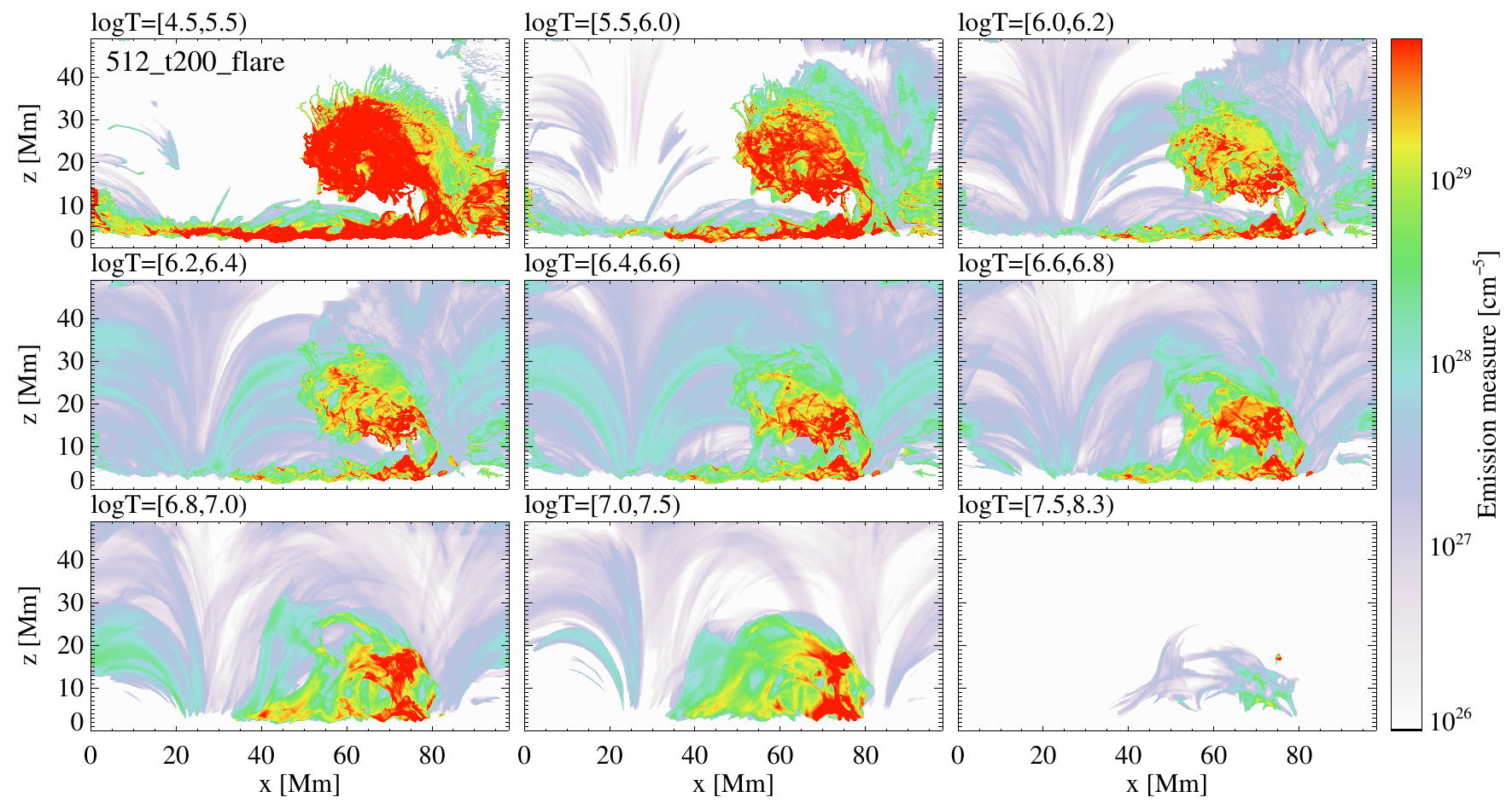}
\caption{EMs (viewed along the $y$-axis) of the ``512\_t200\_flare" run. The snapshot shown corresponds to the time of the peak GOES flux of this run.
\label{fig:em-full-512-t200-flare}} 
\end{figure*}

\begin{figure*}
\includegraphics{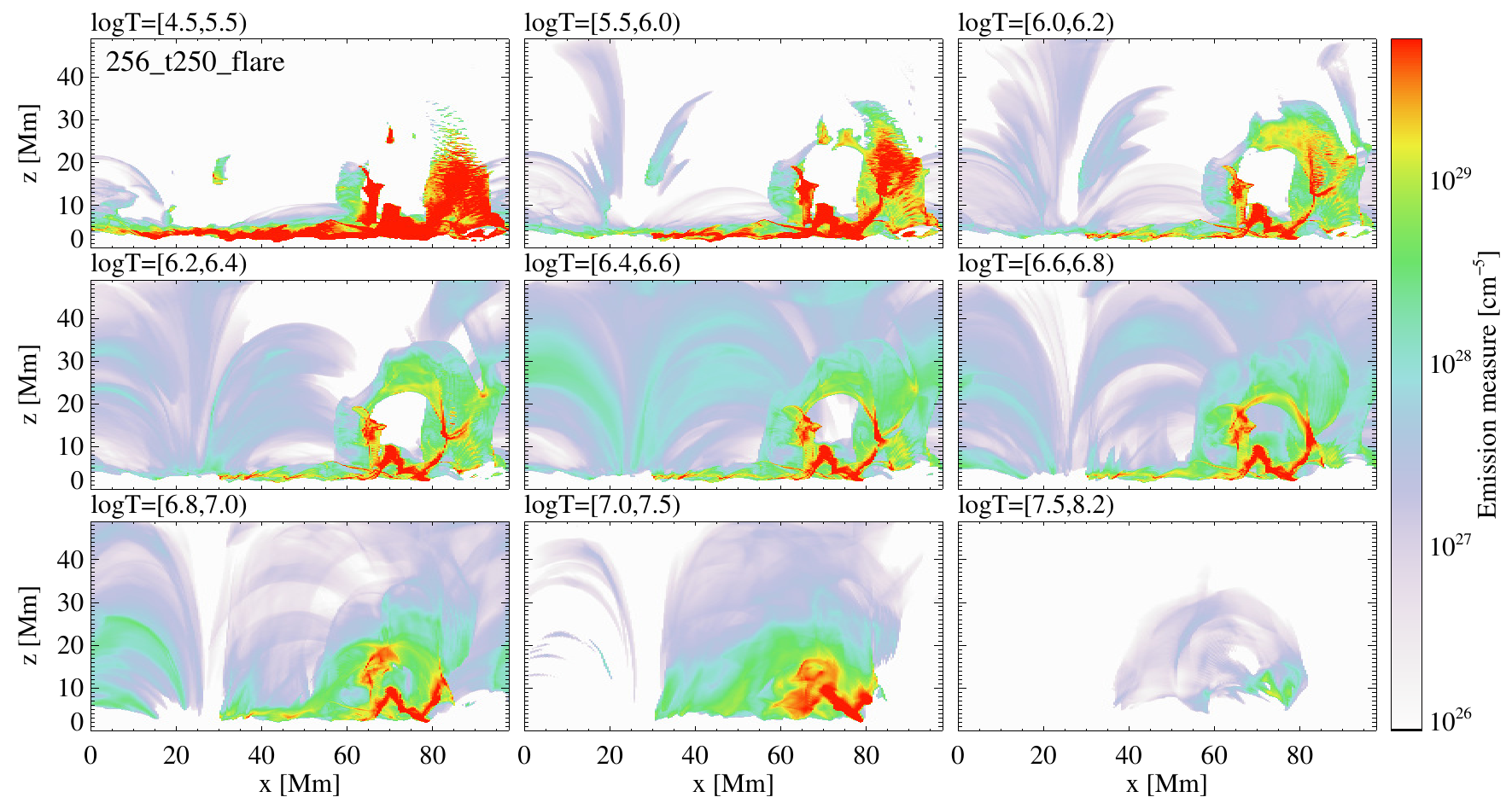}
\caption{EMs (viewed along the $y$-axis) of the ``256\_t250\_flare" run. The snapshot shown is at 1\,min before the time of the peak GOES flux of this run.
\label{fig:em-full-256-t250-flare}} 
\end{figure*}

\begin{figure*}
\includegraphics{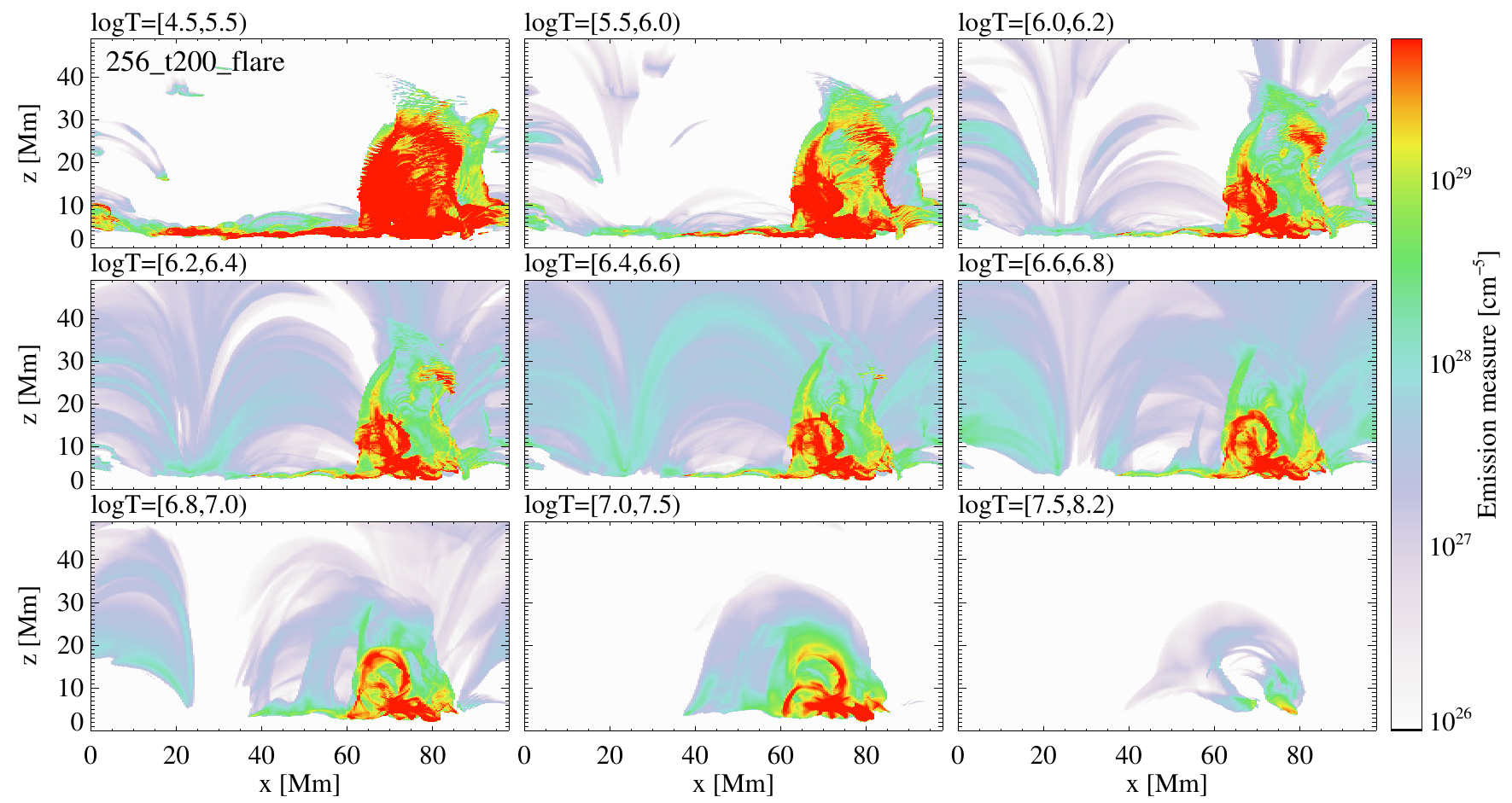}
\caption{EMs (viewed along the $y$-axis) of the ``256\_t200\_flare" run. The snapshot shown corresponds to the time of the peak GOES flux of this run.
\label{fig:em-full-256-t200-flare}} 
\end{figure*}

\begin{figure*}
\includegraphics{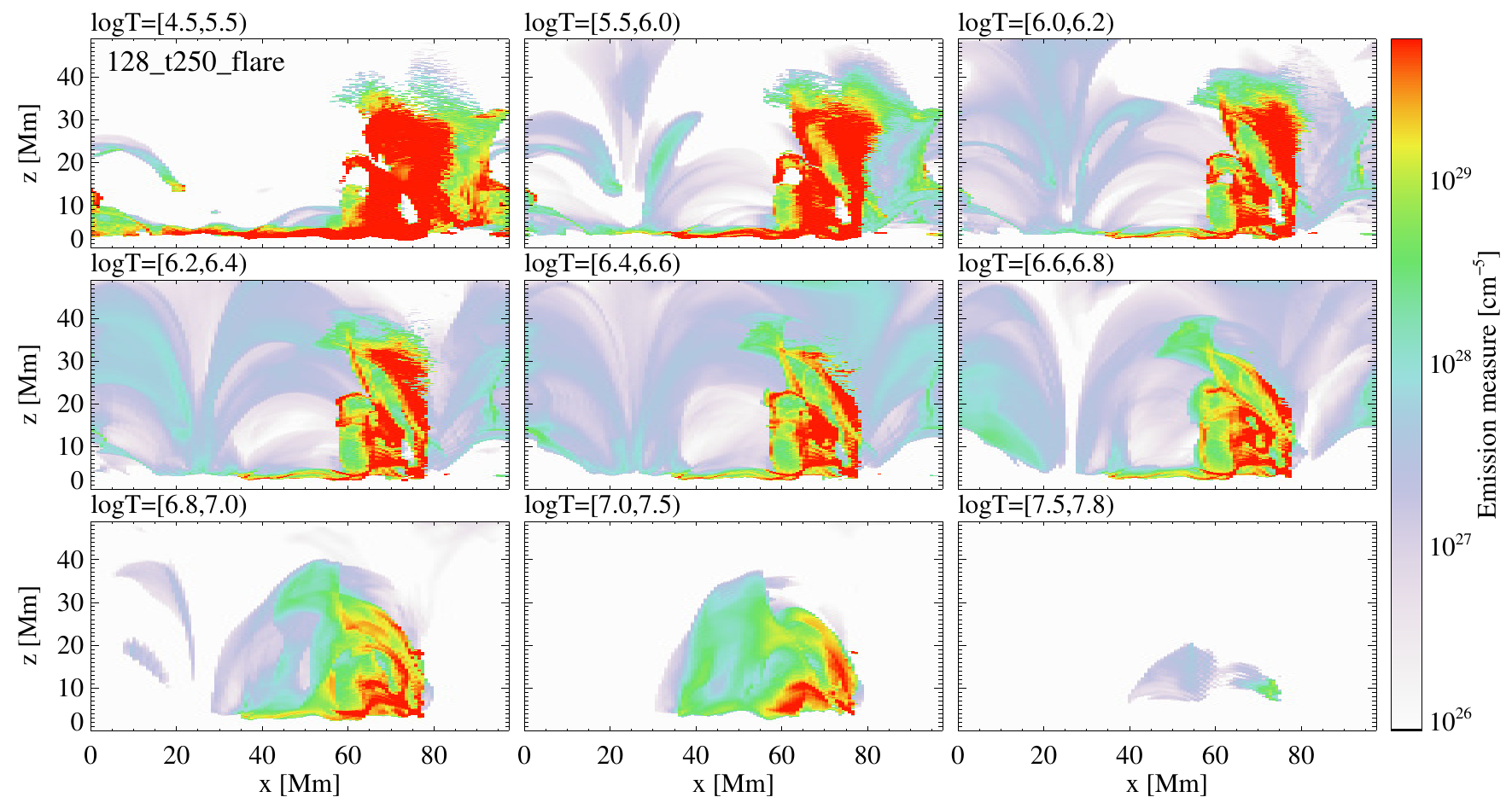}
\caption{EMs (viewed along the $y$-axis) of the ``128\_t250\_flare" run. The snapshot shown is at 40\,s before the time of the peak GOES flux of this run.
\label{fig:em-full-128-t250-flare}} 
\end{figure*}

\begin{figure*}
\includegraphics{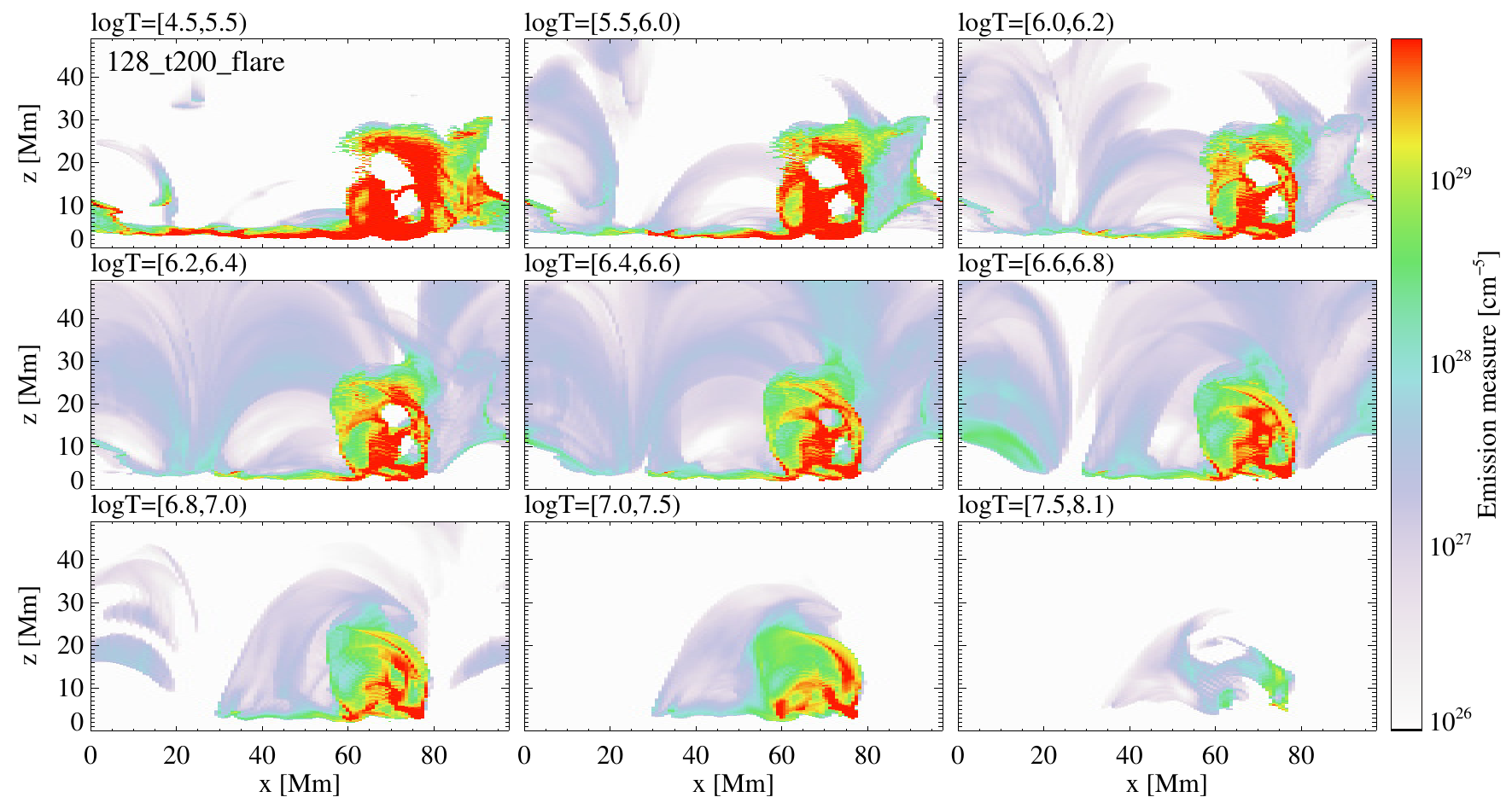}
\caption{EMs (viewed along the $y$-axis) of the ``128\_t200\_flare" run. The snapshot shown is at 9\,s before the time of the peak GOES flux of this run.
\label{fig:em-full-128-t200-flare}} 
\end{figure*}

\begin{figure*}
\includegraphics{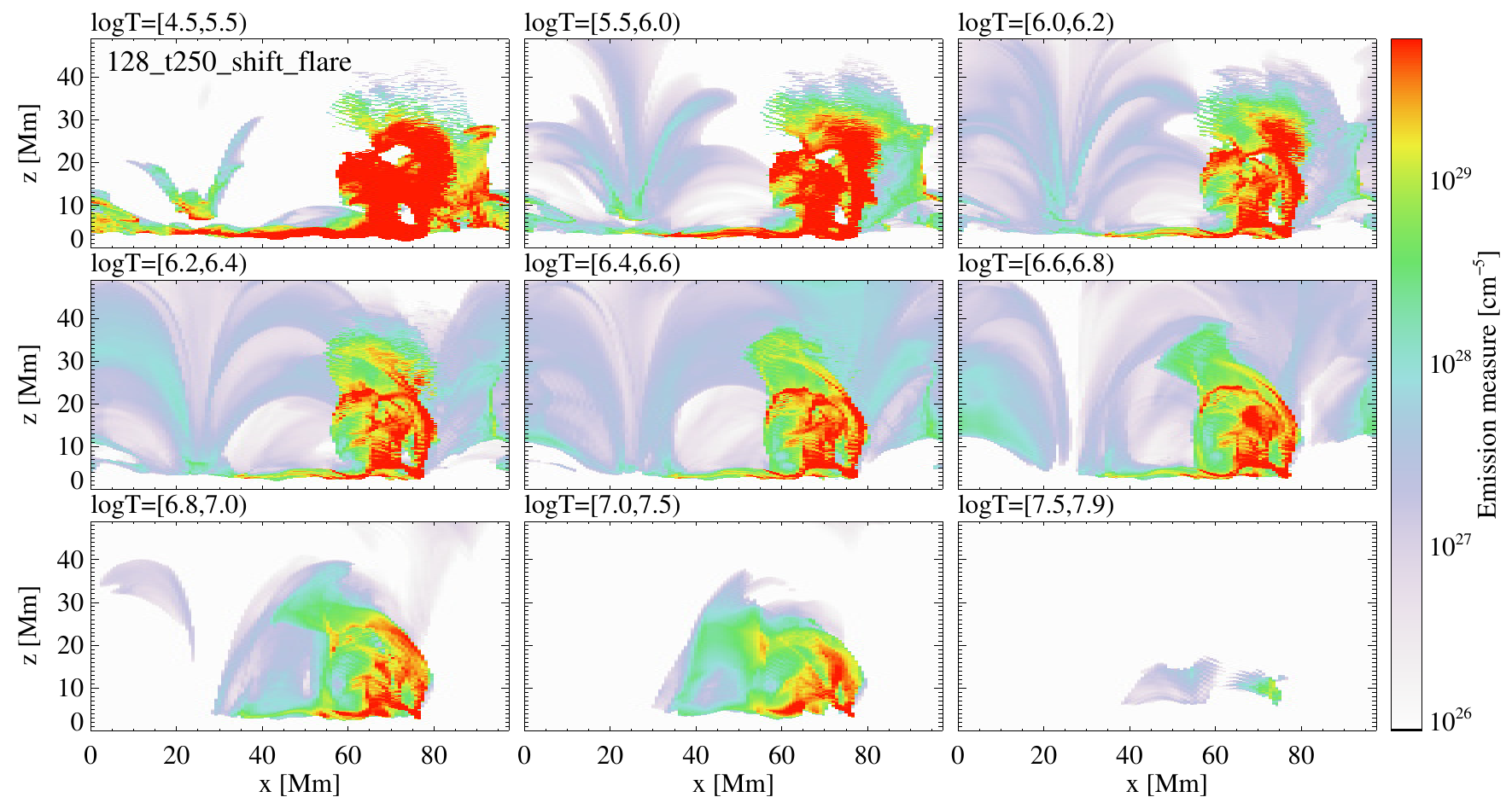}
\caption{EMs (viewed along the $y$-axis) of the ``128\_t250\_shift\_flare" run. The snapshot shown is at 40\,s before the time of the peak GOES flux of this run.
\label{fig:em-full-128-t250-shift-flare}} 
\end{figure*}

\begin{figure*}
\includegraphics{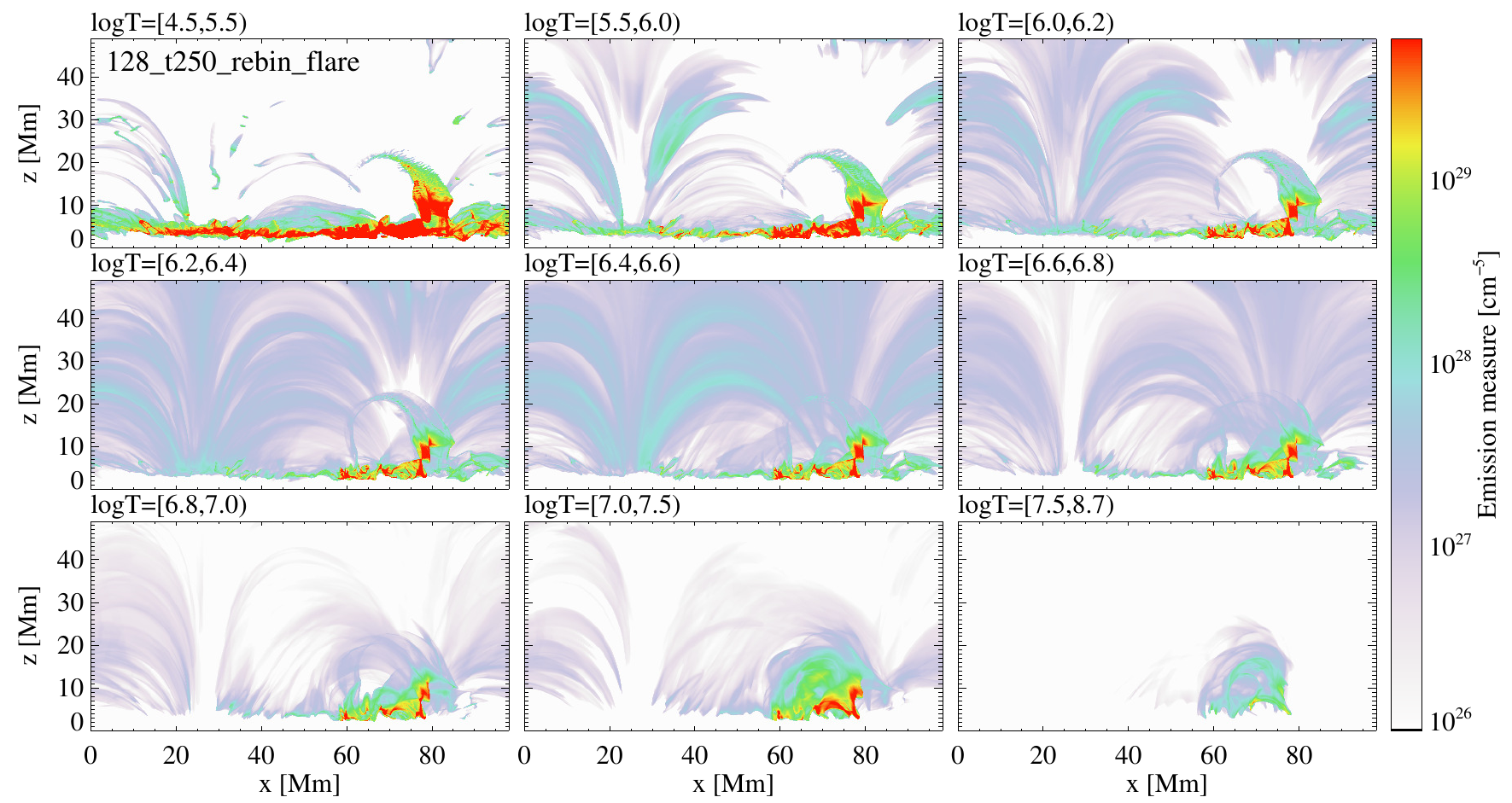}
\caption{EMs (viewed along the $y$-axis) of the ``128\_t250\_shift\_flare" run. The snapshot shown is at 40\,s before the time of the peak GOES flux of this run.
\label{fig:em-full-128-t250-rebin-flare}} 
\end{figure*}

We present in Figures \ref{fig:em-full-512-t250-v3000} -- \ref{fig:em-full-128-t250-rebin-flare} the EMs of the flare runs in the control experiments. As in \figref{fig:em_full}, the original EMs are rebinned to 9 intervals, and all figures are shown in exactly the same colorscale. We note that the detailed temporal evolution of the flares that occur in the control experiments are slightly different; thus, the snapshot displayed for each case intends to represent a similar evolution stage in the sense that the emission structures appear similar to the ground truth.

\end{appendix}
\end{document}